\documentclass[graybox]{svmult}

\usepackage{amsmath}
\usepackage{microtype}
\DisableLigatures{encoding = *, family = * }
\usepackage[hidelinks]{hyperref}
\DeclareMathOperator{\cdf}{cdf}
\DeclareMathOperator{\erfc}{erfc}
\usepackage{tabulary}
\usepackage{booktabs}
\usepackage{helvet}         
\usepackage{courier}        
\usepackage{type1cm}        
\usepackage{makeidx}         
\usepackage{graphicx}        
\graphicspath{{Figures/}}
\usepackage{multicol}        
\usepackage[bottom]{footmisc}

\makeindex             

\begin{document}

\title*{Spatial Stochastic Modeling with MCell and CellBlender}
\authorrunning{Gupta et al.}
\author{Sanjana Gupta, Jacob Czech, Robert Kuczewski, Thomas M. Bartol, Terrence J.
Sejnowski, Robin E. C. Lee, and James R. Faeder}
\institute{Sanjana Gupta, \at Department of Computational and Systems Biology, School of Medicine, University of
Pittsburgh, Pittsburgh PA 15260 USA, \email{sag134@pitt.edu}
\and Jacob Czech \at Pittsburgh Supercomputing Center, Carnegie Mellon University, Pittsburgh, PA 15213
USA.\and Robert Kuczewski \at Howard Hughes Medical Institute, The Salk Institute for Biological Studies, La Jolla,
California 92037, USA \and Thomas M. Bartol \at Howard Hughes Medical Institute, The Salk Institute for Biological Studies, La Jolla,
California 92037, USA\and Terrence J. Sejnowski \at Howard Hughes Medical Institute, The Salk Institute for Biological Studies, La Jolla,
California 92037, USA \and Robin E.C. Lee, \at Department of Computational and Systems Biology, School of Medicine, University of
Pittsburgh, Pittsburgh PA 15260 USA \and James R. Faeder, \at Department of Computational and Systems Biology, School of Medicine, University of
Pittsburgh, Pittsburgh PA 15260 USA, \email{faeder@pitt.edu}}
%
%
\maketitle

\abstract{This chapter provides a brief introduction to the theory and practice of spatial stochastic simulations. It begins with an overview of different methods available for biochemical simulations highlighting their strengths and limitations. Spatial stochastic modeling approaches are indicated when diffusion is relatively slow and spatial inhomogeneities involve relatively small numbers of particles. The popular software package MCell allows particle-based stochastic simulations of biochemical systems in complex three dimensional (3D) geometries, which are important for many cell biology applications. Here, we provide an overview of the simulation algorithms used by MCell and the underlying theory. We then give a tutorial on building and simulating MCell models using the CellBlender graphical user interface, that is built as a plug-in to Blender, a widely-used and freely available software platform for 3D modeling. The tutorial starts with simple models that demonstrate basic MCell functionality and then advances to a number of more complex examples that demonstrate a range of features and provide examples of important biophysical effects that require spatially-resolved stochastic dynamics to capture. \newline This is a preprint of Chapter 24 from Munsky et al., Quantitative Biology: Theory, Computational Methods, and Models (The MIT Press)}

\section{Introduction: Why stochastic spatial modeling?}
\label{sec:1}
Mathematical and computational models are useful tools that can be used to make inferences about experimental data as well as predictions about a system of interest \cite{Goldstein2004,Kholodenko2006}. As outlined in Table~\ref{table:1}, there are many different modeling formalisms available with varying degrees of spatial and molecular resolution.
The degree of spatial resolution roughly divides these methods into two categories: 
\begin{enumerate}
\item Non-spatial --- ordinary differential equations (ODE), stochastic simulation algorithms (SSA), network-free simulations (NF); \index{ordinary differential equations} \index{stochastic simulation algorithm} \index{network-free simulation}
\item Spatial --- partial differential equations (PDE), reaction diffusion master equations (RDME), and particle-based stochastic simulations. \index{partial differential equations} \index{reaction diffusion master equation} \index{particle-based stochastic simulation}
\end{enumerate}
The well-mixed assumption posits that each reacting species is homogeneously distributed throughout each cellular compartment being modeled. The typical distance that a molecule travels over its lifetime, called the Kuramoto length \cite{Grima2008}, can be used to assess the applicability of the well-mixed assumption. The Kuramoto length, $l_k=\sqrt{D \tau}$, where $D$ and $\tau$ are the diffusion coefficient and average lifetime of a given species respectively, is compared with the length scale of the reaction volume, $L$:
\begin{enumerate}
\item If $l_k \gg L$ then local changes in concentration diffuse throughout the reaction volume and the system remains homogeneous/well-mixed.
\item If  $l_k \ll L$ then concentration changes remain localized and the well-mixed assumption breaks down, requiring application of spatial modeling methods.
\end{enumerate}
The degree of molecular resolution, i.e., whether the molecular species are modeled as continuous populations, discrete populations, or discrete individuals, further differentiates the modeling methods, as shown in the second row of Table 1. We now describe these methods in more detail, beginning with the non-spatial methods and then proceeding to spatially-resolved methods, which are required when the well-mixed assumption no longer holds.
\index{Kuramoto length} 
\begin{table}[ht]
\caption{Classification of different methods for biochemical simulation, with examples of available simulation tools.}
\scriptsize
\begin{tabulary}{\linewidth}{LLLLLLL}
\toprule
\textbf{Method} & \textbf{ODE}$^a$ & \textbf{SSA}$^b$ & \textbf{NF}$^c$ & \textbf{PDE}$^d$ & \textbf{Spatial SSA}$^e$ & \textbf{Particle-based}$^f$ 
\\ 
\textit{Attribute} \\
\midrule 
\textit{Spatial resolution} & Well-mixed & Well-mixed & Well-mixed & Lattice & Lattice & Continuous  \\ [5ex]
\textit{Molecular resolution} & Continuous populations & Discrete populations & Discrete individuals & Continuous populations & Discrete populations & Discrete individuals \\ [1ex] \bottomrule 
\end{tabulary}
$^a$Ordinary differential equations \cite{Tyson2003,Aldridge2006}.
$^b$Stochastic simulation algorithm \cite{Gillespie1977}.
$^c$Network-free simulation algorithm, e.g., NFsim \cite{Sneddon2011}
$^d$Partial differential equations, e.g., VCell \cite{Loew2001}.
$^e$Spatial stochastic simulation algorithm, e.g., MesoRD \cite{Elf2004} and URDME \cite{Drawert2012}.
$^f$Particle-based spatial stochastic simulation algorithm, e.g., MCell \cite{Kerr2008} and Smoldyn \cite{Andrews2004}.
\label{table:1} 
\end{table}
\subsection{Non-spatial modeling formalisms}

\textbf{\textit{Ordinary differential equations}} (ODEs) based on reaction rate equations model continuous populations in a well-mixed system \cite{Tyson2003,Aldridge2006}. These deterministic equations track the mean values of the molecular species populations, neglecting the stochastic fluctuations that occur when the number of molecules is small or the system has multiple steady states \cite{Grima2008}. 
\index{ordinary differential equations}
The \textbf{\textit{chemical master equation}} (CME) is a set of coupled linear ODEs that describe the probability distributions of a set of molecular species populations. The CME is used when stochastic effects are expected to be important, such as when either of the following is true:
\index{chemical master equation}
\begin{enumerate}
\item The average number of particles $n$ is small enough that the size of the fluctuations about the mean, which are order $\sqrt{n}$, become significant \cite{Grima2008}.
\item The system has multiple steady state attractors that can give rise to noise-induced transitions between states, which may be either fixed or oscillating \cite{Ullah2010}. 
\end{enumerate}
The CME can only be solved directly for systems with a small number of degrees of freedom, but \textit{system trajectories} can be computed using various methods including Gillespie's \textbf{\textit{stochastic simulation algorithm}} (SSA) \cite{Gillespie1977}, and then binned or averaged to determine probability distributions and moments. Because the CME is a population-based method, it only permits tracking the collective behavior of species in the system. If the modeler is interested in tracking individual species, \textbf{\textit{network-free}} (NF) simulation approaches can be used \cite{Sneddon2011,Chylek2013}. These individual-based Monte-Carlo based algorithms have an additional advantage in that they avoid enumeration of the complete reaction network in advance, which can be prohibitively expensive for systems that exhibit combinatorial complexity \cite{Sneddon2011}. 
\index{combinatorial complexity}

\subsection{Spatial modeling formalisms}

There are many examples in biology of systems where the well-mixed assumption does not hold throughout the entire reaction volume, requiring a different set of modeling formalisms that take into account spatial effects. 

\textbf{\textit{Partial differential equations}} (PDEs) are used to model continuous molecular populations as they evolve in time and space. For simple systems, these can be solved analytically with a continuous representation of space \cite{Lipkow2008,Tindall2008,Luckhaus1992}. However, most complex systems rely on numerical integration methods that discretize space \cite{Loew2001}. As in the case of ODEs, PDEs are deterministic and break down when stochastic effects are important. An example of this is the Min system in Escherichia coli. Low copy numbers of proteins in this system can produce spatial patterning that does not occur in the corresponding deterministic model \cite{Howard2003}. Under some conditions, this system exhibits both a stable fixed point and a limit cycle attractor. Stochastic fluctuations drive the system into a limit cycle and render the stable fixed point unimportant for the observed dynamics \cite{Fange2006}.
\index{Min system}

The \textbf{\textit{reaction diffusion master equation}} (RDME), which is the spatial extension of CME, divides the space into sub-volumes called voxels, within which well-mixed conditions are assumed to prevail \cite{Gillespie2014,Gillespie2013}. Reactions occurring within individual voxels are simulated using Gillespie's algorithm, and molecules move from one voxel to a neighboring one via discrete jumps to simulate diffusion. The set of voxels used to model a geometry is called a mesh: determining the geometry and resolution of the mesh is generally the most difficult part of performing RDME-based simulations. Rectangular mesh elements are appropriate for modeling relatively simple geometries, but the more complex geometries that are often encountered in realistic simulations of cells require so-called unstructured meshes, based on triangles or tetrahedrons. Software packages are available for both rectangular \cite{Hattne2005} and unstructured mesh elements \cite{Drawert2012}.
Optimal determination of mesh resolution is important because the computational cost of these methods grows rapidly as the mesh size increases due to the increase in the number of diffusive jump events relative to the number of reaction events, but is unfortunately largely a matter of trial and error \cite{Gillespie2013}.
\index{reaction diffusion master equation}

\textbf{\textit{Particle-based spatial stochastic simulation methods}} simulate the trajectories of individual particles that represent the molecular species in the system. In addition to allowing the modeler to track individual particles, so-called off-lattice methods that form the basis for the popular MCell \cite{Kerr2008,Bartol2014} and Smoldyn \cite{Andrews2010} simulators, do not rely on discretization of three-dimensional (3D) space and can therefore circumvent the complexities involved in mesh generation mentioned above. At the same time, these tools enable modeling of complex 3D geometries that often arise in cell biology applications, such as studying the effects of macromolecular crowding, oligomerization, and self-assembly \cite{Schoneberg2014}.
\index{particle-based stochastic simulation} \index{molecular crowding}

\section{A brief overview of MCell}

MCell, which is short for Monte Carlo Cell, is a particle-based reaction diffusion simulator \cite{Kerr2008,Bartol2014,Bartol1991,Stiles1996,Stiles2001}. A detailed description of the simulation algorithms used in MCell3, the current version, is presented in Kerr et al. \cite{Kerr2008}. Here, we describe the basic components of an MCell model and the basic elements of an MCell simulation step. An MCell simulation consists of a trajectory generated by a specified number of steps of a fixed time length. 

\subsection{Basic components}
\label{sec:components}
\begin{figure}[tbh]
\centering
\includegraphics[width=\textwidth]{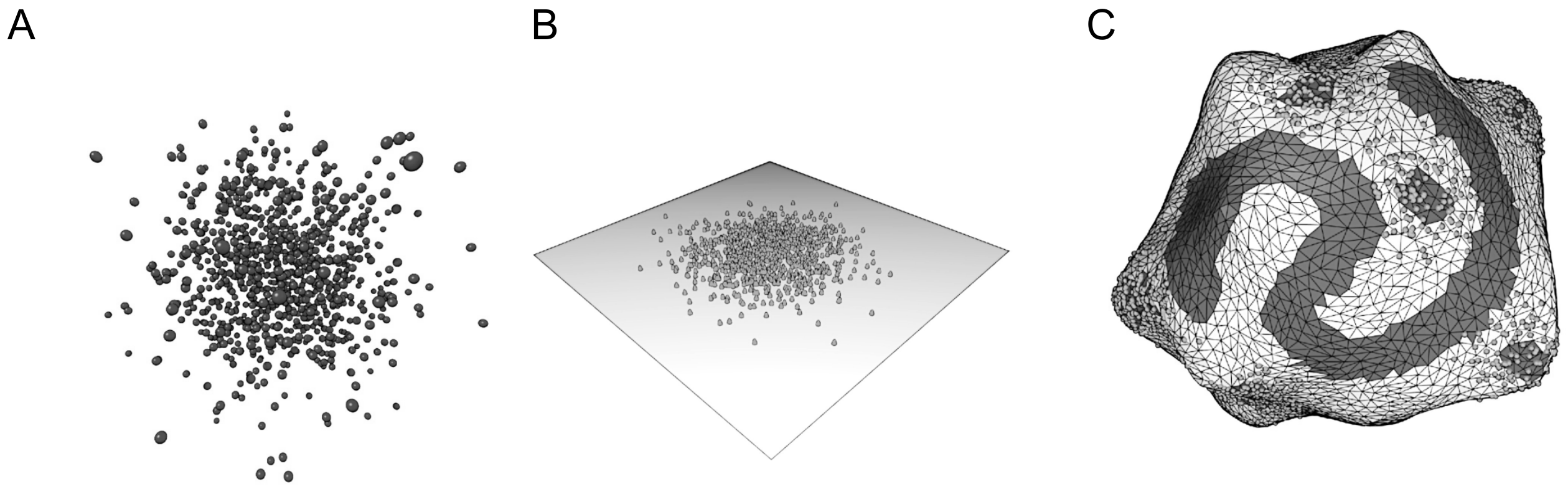}
\caption{MCell Components. (A) Volume Molecules diffusing in free space. (B) Mesh Object defined by a Plane with Surface Molecules diffusing on it. (C) Mesh Object defined by a complex closed mesh with multiple defined Surface Regions, in which Surface Molecules have different diffusion constants, as defined by corresponding Surface Classes.}
\label{figure:1}
\end{figure}

The basic components of an MCell model are:
\begin{enumerate}
\item \textbf{Molecules.}  Molecules are the fundamental components of the system and are simulated as diffusing point particles. Surface Molecules diffuse on 2D surfaces and Volume Molecules diffuse in 3D volume (Figure~\ref{figure:1}(A)). 
\item \textbf{Reactions.} Reactions define unimolecular or bimolecular reactions with mass action kinetics that determine how the molecules in the model interact with each other. These are discussed in more detail below.
\item \textbf{Mesh Objects.} Mesh Objects, or ``Objects'' for short, define surfaces that limit the diffusion of associated Surface Molecules and may reflect, absorb, or transmit colliding Volume Molecules. Objects may be open (Figure~\ref{figure:1}(B)), but are more commonly closed in order to form compartments that restrict the diffusion of Volume Molecules (Figure~\ref{figure:1}(C)).
Objects are composed of triangular faces, which allows the creation of complex curved geometries. Objects may be divided into Surface Regions, comprising a set of faces, that may affect molecule behavior and placement (Figure~\ref{figure:1}(C)). Properties of Surface Regions are assigned using Surface Classes. It is worth noting that within the MCell simulator, faces defining an Object are further divided into triangular tiles, each of which may only be occupied by a single Surface Molecule. 
\item \textbf{Release sites.} Release Sites define where molecules are placed at the start of a simulation or during a simulation. Examples of Release Sites are the surface of an Object, the interior or exterior of an Object, points in space, and predefined geometric shapes. 
\end{enumerate}
\index{mesh}
\index{diffusion}

\subsection{Simulation algorithm}
\begin{figure}[tbh]
\centering
\includegraphics[width=0.5\textwidth]{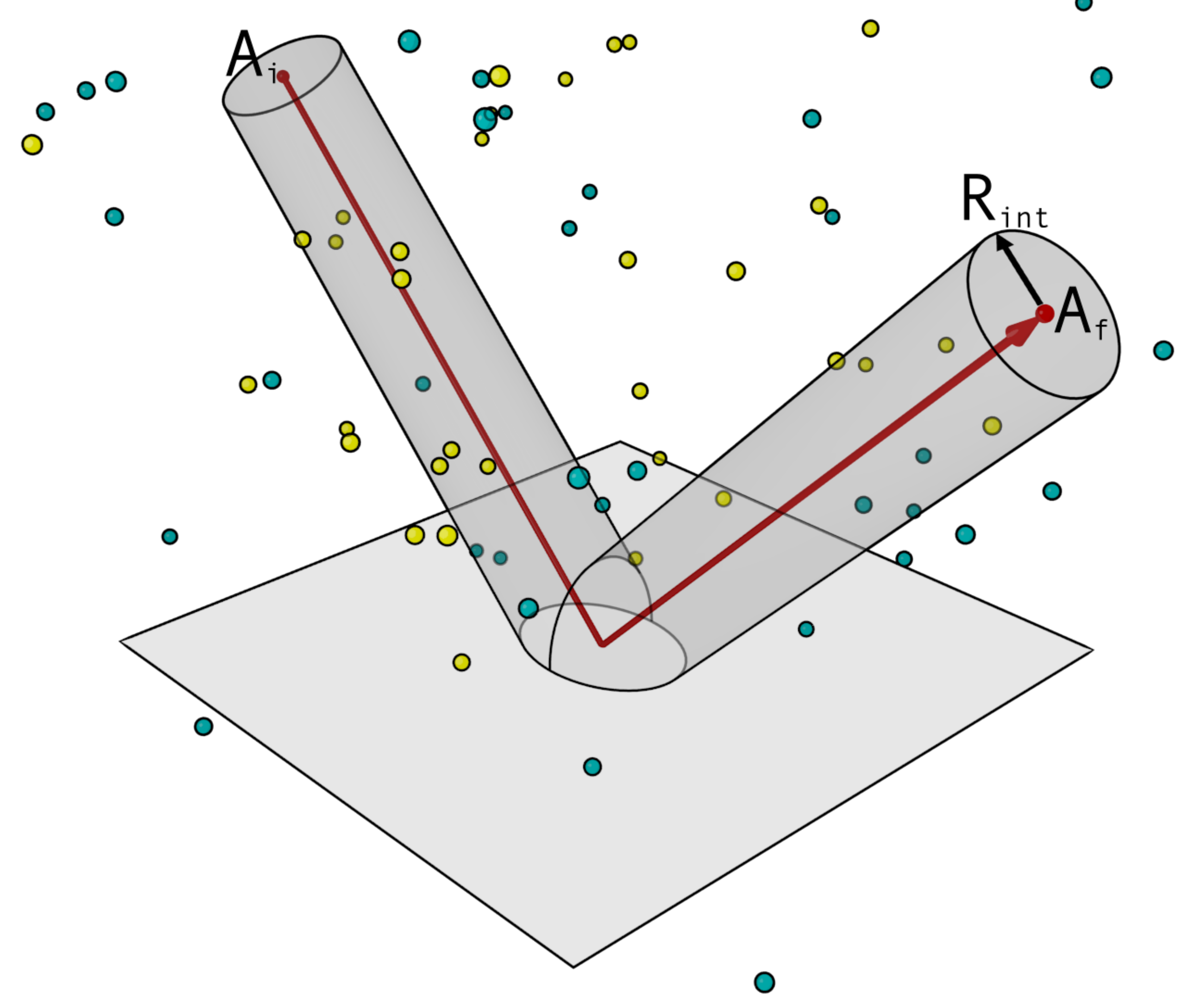}
\caption{Simulation of a Volume Molecule time step in MCell. The particle A diffuses along a straight line path of a pre-determined length and is tested for reaction partners within a specified interaction radius, $R_\text{int}$, along its path. When a particle encounters a mesh tile, it is reflected according to specular reflection and continues in a straight line.
}
\label{figure:2}
\end{figure}

At every time step in an MCell simulation, each particle can move, collide with other particles or surfaces, and undergo bimolecular and unimolecular reactions. The basic elements of a simulation step (Figure~\ref{figure:2}) are:	
\begin{enumerate}
\item \textbf{Movement}. Particles move in random directions over straight-line trajectories of length sampled from probability distributions for diffusive motion, as described in more detail in Section~\ref{section:diffusion}.
\item \textbf{Detection of collisions}. Ray-tracing algorithms detect collisions between the diffusing particles and other particles or surfaces as it diffuses. As a particle moves along a straight-line trajectory, any particle within a specified collision radius is checked to determine if the pair undergoes a reaction (Figure~\ref{figure:2}). Ray marching algorithms propagate rays after collisions with surfaces.
\item \textbf{Bimolecular reactions}. When a collision occurs, the bimolecular reaction probability is a function of the user-specified bimolecular rate constant, the time step, and the diffusion constants of the particles. A Monte Carlo scheme is used to determine reaction firings, as described further in Section~\ref{section:bimolecular}. 
\item \textbf{Unimolecular reactions}. At any point along its trajectory a particle can undergo unimolecular transitions based on user-specified reactions. These transition events are scheduled whenever a particle is created or changes state and are determined probabilistically as a function of the rate constants for the allowed reactions, as described in more detail in Section~\ref{section:unimolecular}.
\end{enumerate}
\index{ray tracing}
\index{bimolecular reaction}
\index{unimolecular reaction}

\subsection{Particle diffusion}
\label{section:diffusion}
Following \cite{Kerr2008} we briefly describe the steps to simulate particle diffusion and the equations needed to determine both the magnitude and direction of particle displacement at each time step. 
Given a Volume Molecule at a point in 3D space at time $t=0$, the probability density function for it having moved a distance $r$ by time $t$ is given by
\begin{equation}
\rho(r,t) = \frac{1}{(4\pi D t)^{3/2}}e^{-r^2/4D t}.
\end{equation}
Inverse transform sampling is used to sample a displacement magnitude $R$ from this distribution by computing 
$R=\cdf^{-1} X$, where $X$ is a uniform random number between 0 and 1 and  
\begin{equation}
\cdf(R,t)= \int_0^R \rho(r,t) \;4\pi r^2 \;\mathrm{d}r.
\label{eq:cdf}
\end{equation}
This function can be expressed in terms of the Gaussian error function and its inverse computed from a lookup table \cite{Kerr2008}. 
\index{diffusion} \index{diffusion equation}

In addition to the displacement magnitude, two angles are needed to determine the direction of displacement: the azimuthal angle $\phi$  and the polar angle $\theta$.  $\phi$ is chosen randomly from the uniform distribution on the interval $\left[0,2\pi\right)$. $\theta$ is more complicated because the probability distribution is proportional to $\sin{\theta}$. Inverse transform sampling can be used to obtain $\theta$ from the equation $Y=(1-\cos{\theta})/2$, where $Y$ is a uniform random number on $\left[0,1\right]$. Since computing the inverse of a trigonometric function is computationally expensive, a modified form of table lookup is used to boost efficiency \cite{Kerr2008}.

Overall, the steps for simulating Volume Molecule diffusion can be summarized as follows:
\begin{enumerate}
\item Determine the radial displacement  $R$ using inverse transform sampling of the $\cdf{}$ given in Equation~\ref{eq:cdf}.
\item Determine the direction of travel by sampling the polar angles $\theta$ and $\phi$.
\end{enumerate}

MCell also considers Surface Molecules that diffuse in 2D. As described in Section~\ref{sec:components}, Mesh Objects are composed of triangular Faces, which in turn are divided into smaller triangles called Tiles.  Surface Molecules diffuse by hopping between tiles. Unlike Volume Molecules, which are treated as point particles and can exist at arbitrarily close points in space, Surface Molecules are restricted from occupying a Tile already occupied by another Surface Molecule. The algorithm for simulating 2D diffusion in MCell is similar to algorithms for simulating the Reaction Diffusion Master Equation \cite{Gillespie2013} and are described in more detail in \cite{Kerr2008}. The basic steps can be summarized as follows:
\begin{enumerate}
\item Pick a direction of motion within the plane of the Tile on which the Molecule currently resides.
\item Ray-march along this vector.
\item As the ray crosses triangle boundaries, convert the vector from the local planar coordinate system of the first triangle to that of the second.
\end{enumerate}
\index{surface diffusion}

\subsection{Bimolecular reactions}
\label{section:bimolecular}
MCell checks for the occurrence of a bimolecular reaction whenever two particles collide. To obtain the correct bulk reaction rate, the probability of reaction per collision, $p$, must be chosen such the expected rate of collisions times the probability per collision equals the bulk reaction rate. We will briefly derive the reaction probability for the case of a Surface Molecule colliding with a Volume Molecule, and provide results for the surface-surface and volume-volume cases, which are derived in detail elsewhere \cite{Kerr2008}.

\begin{figure}[tbh]
\centering
\includegraphics[width=0.5\textwidth]{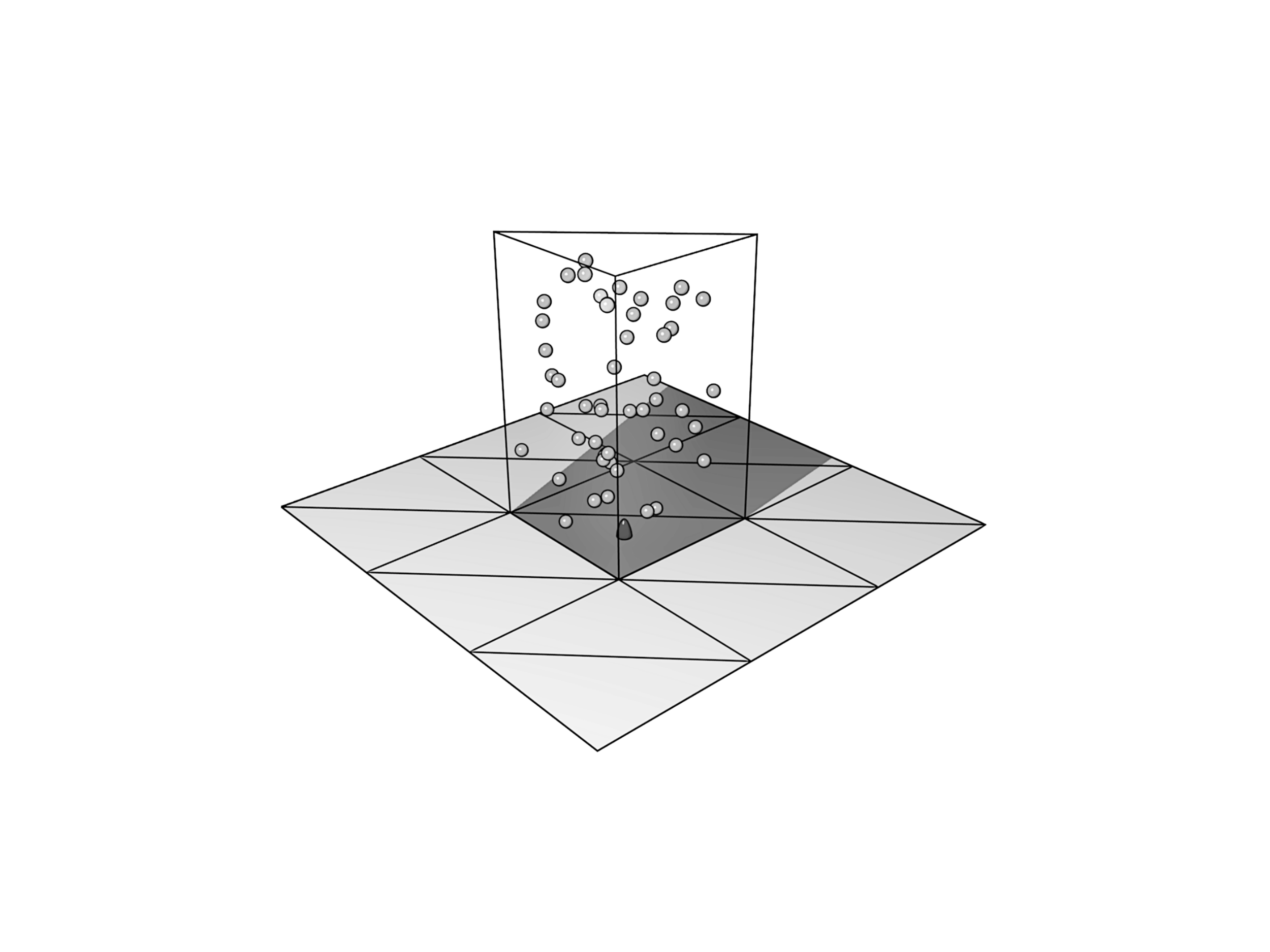}
\caption{Surface-volume reactions. A Surface Molecule (dark protrusion) on a Tile of area $A$ with a column directly above containing Volume Molecules (shaded spheres).}
\label{figure:3}
\end{figure}

\textbf{Surface--volume reactions}.
Consider a Surface Molecule located on a tile of area $A$ with an infinite column extending above (Figure~\ref{figure:3}). At equilibrium, the flux of molecules moving into this column equals the flux of molecules moving out of this column, and therefore we can restrict ourselves to the Volume Molecules within the column. The probability that a molecule starting at distance $R$ away will reach the surface is
\begin{equation}
p_s(R)= \int_R^\infty \rho(x,\Delta t) \;\mathrm{d}x= \int_R^\infty \frac{1}{\sqrt{4\pi D \Delta t}}\; e^{-x^2/4 D \Delta t} \;\mathrm{d}x = \frac{1}{2} \erfc{\frac{R}{\sqrt{4 D \Delta t}}},
\end{equation}
where $\rho(x,\Delta t)$ is the probability density for a molecule to diffuse by a displacement $x$ in time $t$. The number of molecules in a slice of the column of width $\mathrm{d}R$ with a concentration of Volume Molecules $\rho_1$  is $\rho_1 A\;\mathrm{d}R$, which allows us to determine the total number of Volume Molecules hitting the surface per time step as 
\begin{equation}
n_s = \int_0^\infty p_s(R)  \rho_1 A\;\mathrm{d}R = \int_0^\infty \frac{1}{2} \erfc{\frac{R}{\sqrt{4 D \Delta t}}}
\rho_1 A\;\mathrm{d}R= \frac{\rho_1 A \sqrt{D \Delta t}}{\sqrt{\pi}\Delta t}.
\end{equation}
Now, setting the observed reaction rate, which is $n_s\times p$, equal to the bulk reaction rate, $k \rho_1$, allows us to solve for $p$ to obtain
\begin{equation}
p=\frac{k}{A}\left( \frac{\pi \Delta t}{D}\right)^{1/2}.
\end{equation}
We see that the probability of reaction per collision is proportional to the bimolecular rate constant $k$  but the square root of  $\Delta t$. If the reaction probability is greater than one, the reaction rate is considered \emph{diffusion-limited}, because a reaction will occur any time the reacting particles collide and will be less than that specified by the bulk reaction rate constant. Unless one is explicitly trying to model diffusion-limited reactions, a reaction probability above one usually means that the step size needs to be reduced in order to prevent a loss of reaction flux.
\index{bimolecular rate constant}
\index{surface reaction}

\textbf{Surface-surface reactions}.
A similar derivation gives the reaction probability for bimolecular reactions between two Surface Molecules as
\begin{equation}
p= k \frac{\Delta t}{A}.
\end{equation}

\textbf{Volume--volume reactions}.
The reaction probability for bimolecular reactions between two Volume Molecules with diffusion length constants $D_1$ and $D_2$ respectively, is 
\begin{equation}
p= k \frac{\sqrt{\pi \Delta t}}{4A_\text{int}\left(\sqrt{D_1}+ \sqrt{D_2}\right)},
\end{equation}
where $A_\text{int}=\pi r_\text{int}^2$ and $r_\text{int}$ is the collision radius of the two Volume Molecules.

\subsection{Unimolecular reactions}
\label{section:unimolecular}
When a new molecule is created in an MCell simulation, either through the occurrence of a reaction or at the start of a simulation, the time of the next unimolecular reaction that it will undergo is computed and added to a scheduler. The algorithm for determining the time and identity of the next unimolecular reaction is a special case of Gillespie's SSA \cite{Gillespie1977}.
\index{stochastic simulation algorithm}

\section{Getting started with CellBlender and MCell}

We now present a tutorial that describes how to use MCell via CellBlender, which is a plug-in for the popular 3D modeling software called Blender. CellBlender provides a graphical user interface (GUI) for building and simulating MCell models. Blender, CellBlender, and MCell are each freely available and open source software packages. 
\index{Blender}

MCell models can also be built directly without the CellBlender GUI, using the Model Description Language (MDL). CellBlender generates MDL files automatically, making it possible for a user to create and run a model in MCell without learning MDL, but for advanced applications a knowledge of MDL may be useful. For more details about MDL see the MCell Quick Reference Guide and MCell Reaction Syntax documents available at \url{http://mcell.org/documentation}. 

The tutorial presents a series of examples of increasing complexity that cover most major aspects of spatial modeling with CellBlender and MCell. Each example is divided into two parts:  (1) CellBlender preliminaries, which describe how the modeled components are created using the interface, and (2) a modeling exercise, in which the user is instructed to build a specific model and results of the model are presented and discussed. Complete versions of all of the example models can be obtained from the MCell web site.

\subsection{Obtaining the required software}

\begin{figure}[tbh]
\centering
\includegraphics[width=\textwidth]{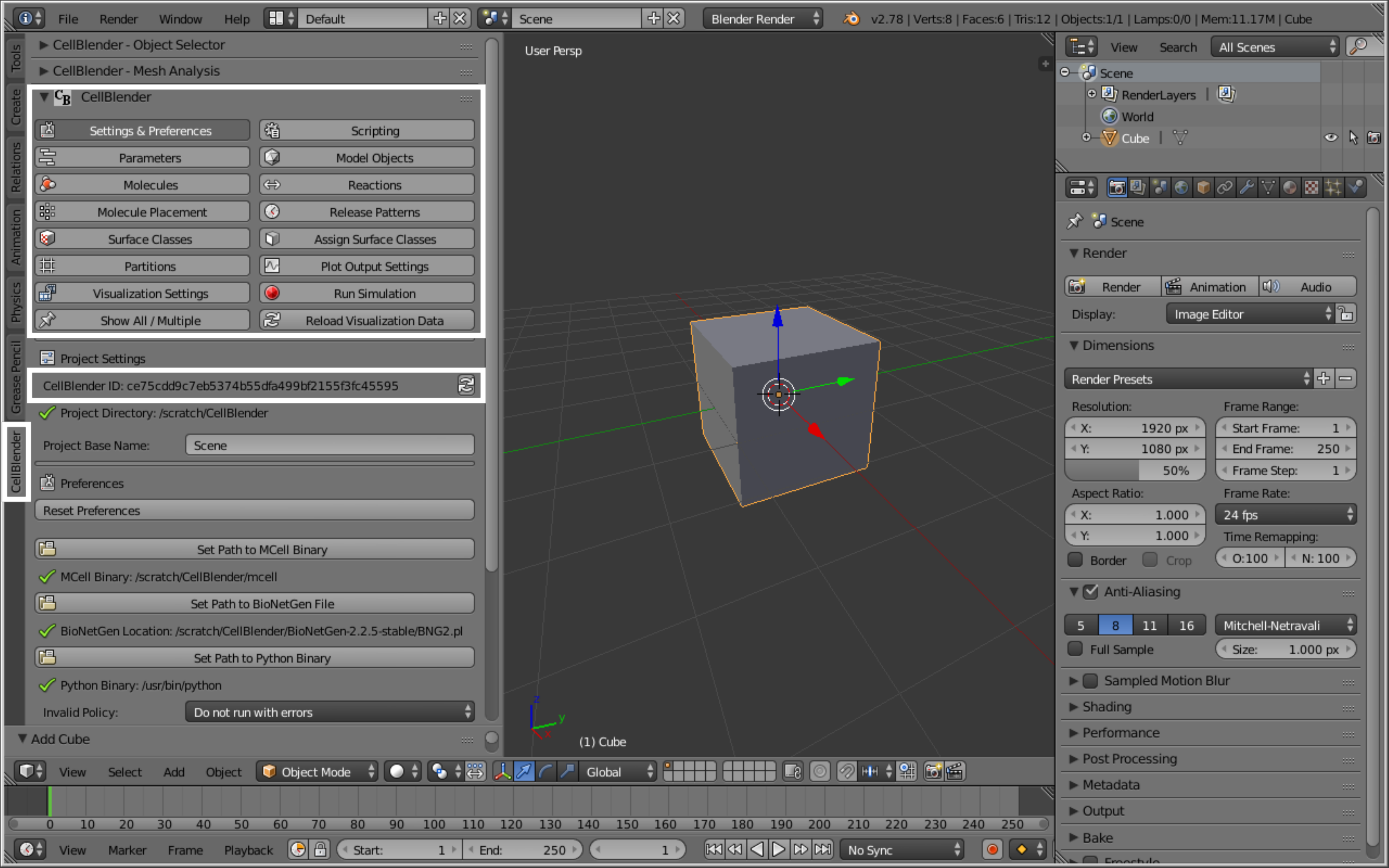}
\caption{The CellBlender interface. Selection of the CellBlender tab in the Blender window (far left), brings up the CellBlender panel which consists of a set of buttons used to define the model. Selecting the \textit{Settings \& Preferences} button displays the \textit{CellBlender ID}, which is useful for to report in publications for replicability and for reporting issues.}
\label{figure:4}
\end{figure}

The first step is to obtain CellBlender and MCell by following the directions at \url{http://www.mcell.org/tutorials/software.html}.
This tutorial is based on CellBlender 1.1, and should be compatible with future version 1 releases. Figure~\ref{figure:4} shows an overview of the CellBlender 1.1 interface.
As shown in Figure~\ref{figure:4}, this tutorial was constructed using CellBlender  with an ID of \texttt{ce75cdd9c7eb5374b55dfa499bf2155f3fc45595}.

Depending on the download configuration used for installation, there may be a number of objects in Blender's central 3D window---typically a cube, a camera, and a lamp. If so, they should be removed prior to starting an exercise. To do this use the mouse to move the pointer into the central 3D window and hit the ``a'' key to select all objects in the window. The selected objects will be outlined in orange. (Pressing the ``a'' key again would toggle the selection, and all objects would have a black outline). Next, hit the ``x'' key, which will bring up a dialog box containing a  \textit{Delete} button that will enable you to delete the selected objects. Select \textit{Save Startup File} from the \textit{File} menu and click the highlighted text to confirm. This will ensure that Blender initiates with an empty window that is best suited for doing the exercises below.
  
\subsection{A brief note about units}

The following story illustrates the pitfalls of not handling units correctly in a model:
\begin{quotation}
Designed to orbit Mars as the first interplanetary weather satellite, the Mars Orbiter was lost in 1999 because the NASA team used metric units while a contractor used imperial. The \$125 million probe came too close to Mars as it tried to maneuver into orbit, and is thought to have been destroyed by the planet's atmosphere. An investigation said the root cause of the loss was the failed translation of English units into metric units in a piece of ground software \cite{BBC2014}.
\end{quotation}

So be warned! Spatial dimensions in MCell are in microns ($1\mu \text{m}=10^{-6}\text{m}$). Time is in seconds (s). Diffusion coefficients are specified in units of $\text{cm}^2/\text{s}$. Molecule amounts can be specified as numbers of molecules or as concentrations in molar (M) units for Volume Molecules and $\mu\text{m}^{-2}$ for Surface Molecules. Unimolecular rate constants are given in $1/\text{s}$.
Bimolecular reaction rate constants between two Volume Molecules or a Volume and Surface Molecule are in $\text{M}^{-1}\text{s}^{-1}$, while bimolecular reaction rate constants between two Surface Molecules are in $\mu\text{m}^{2}\text{s}^{-1}$. Each simulation runs for a specified number of iterations, and the duration of an iteration is one time step.
\index{units}

\section{Simulating free molecular diffusion}

Diffusing particles are a fundamental component of any MCell model. In this section, we consider the case of a single type of molecule freely diffusing in space.

\subsection{CellBlender preliminaries}

\begin{figure}[tbh]
\centering
\includegraphics[width=\textwidth]{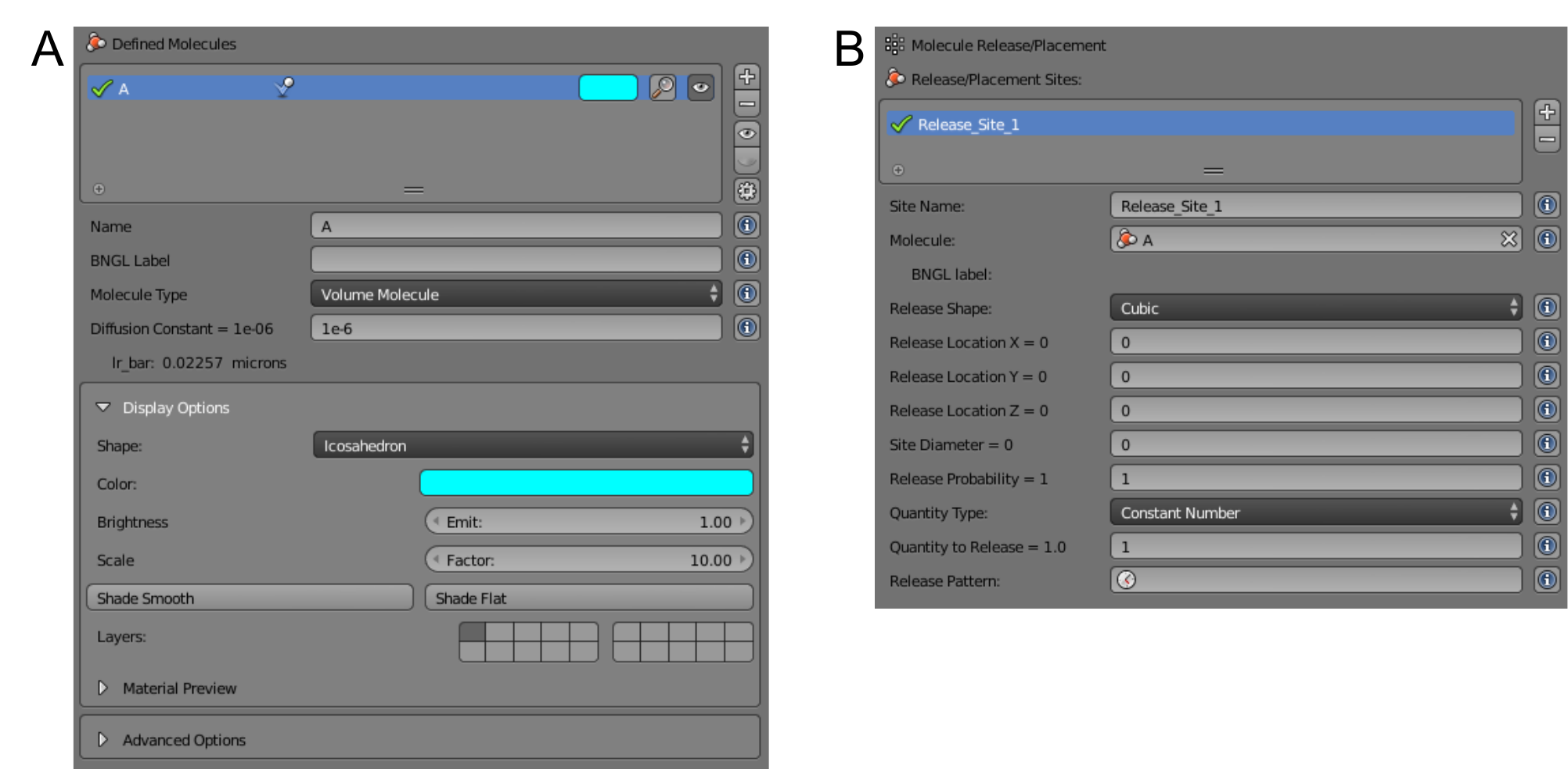}
\caption{CellBlender interface tabs for defining molecules and their Release Sites. (A) The \textit{Define Molecules} panel. (B) The \textit{Molecule Release/Placement} panel. See text for details.}
\label{figure:5}
\end{figure}

To set up this model we will need to make use of the following elements of the CellBlender interface (Figure~\ref{figure:4}).:

\textbf{Molecules}. Figure~\ref{figure:5}(A) shows the \textit{Define Molecules} panel. As in other CellBlender tabs, the plus sign to the right of the \textit{Define Molecules} panel is used to add a new element, in this case a Molecule. Properties of the molecule are specified by entering its Name, Type (Volume or Surface), and Diffusion Constant. The minus sign can be used to delete a previously defined molecule. The \textit{Display Options} control the molecule's shape, size, color, and brightness in the display---note that these visualization options have no effect on the simulation. The quantity ``lr\_bar'' that appears underneath the diffusion constant entry (Figure~\ref{figure:5}(A)) reports the mean diffusion distance given the value of the diffusion constant and the current time step.

\textbf{Molecule Placement}. Molecules are placed into the simulation geometry by defining Release Sites using the \textit{Molecule Release/Placement} panel (Figure~\ref{figure:5}(B)). Fully specifying a Release Site involves defining Site Name, Molecule (selected from the list of defined Molecules), Release Shape (selected from a list), and additional options that depend whether Object/Region or one of the three defined Release Shapes (Spherical Shell, Spherical, or Cubic) is selected. Here, the options required when the Cubic Release Shape is selected are shown. These define the position and size of the Release Shape. The other required items are Quantity Type, usually either Concentration/Density or Constant Number, and Quantity to Release, which will be specified as either a concentration (M) or number of molecules. The Release Pattern, which controls the timing of the release either through explicit timing parameters or triggers defined by reactions, will not be discussed further in the tutorial. 

\begin{figure}[tbh]
\centering
\includegraphics[width=\textwidth]{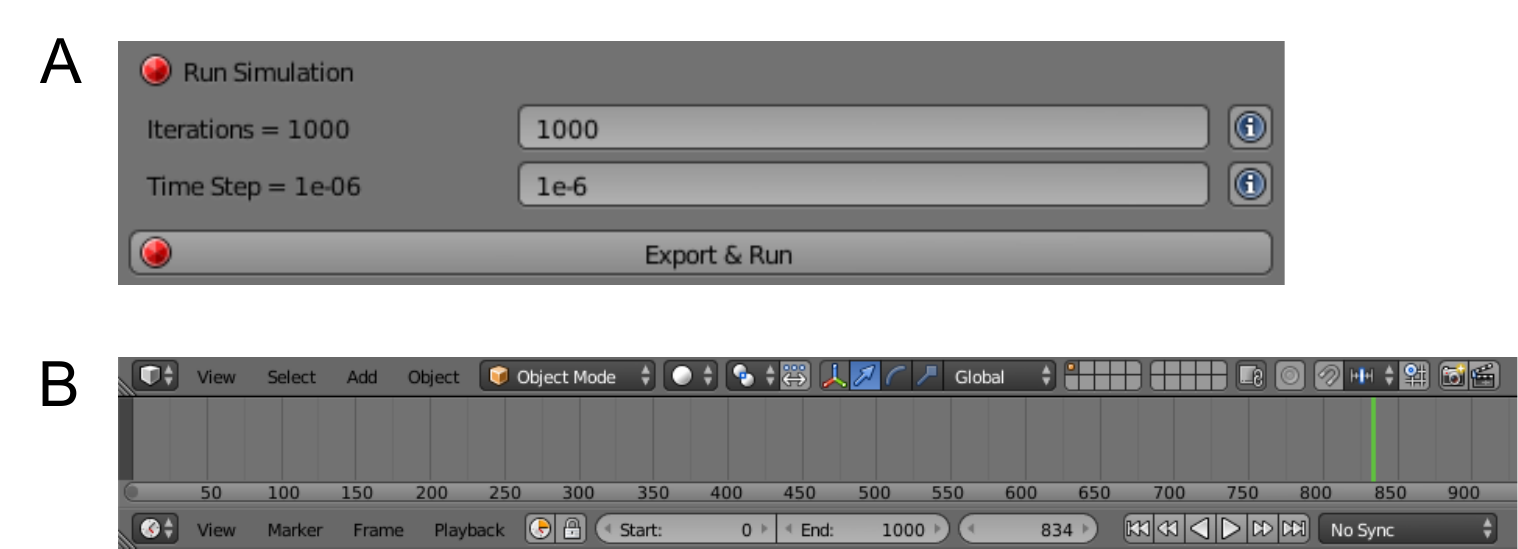}
\caption{Running and visualizing simulations. (A) The \textit{Run Simulation} panel. (B) The Blender time line, which appears at the bottom of the Blender window (see Figure~\ref{figure:4}).}
\label{figure:6}
\end{figure}

\textbf{Run Simulation}. The \textit{Run Simulation} panel (Figure~\ref{figure:6}(A)) is relatively simple. The number of iterations and the time step of each iteration are defined before launching the simulation with the \textit{Run} button. 

\textbf{Reload Visualization data}. The user clicks the \textit{Reload Visualization Data} button to load and view the results of the simulation. The time line at the bottom of the Blender window (Figure~\ref{figure:6}(B)) provides controls to play, pause, and scroll through the simulation.

\subsection{Exercise: simple diffusion}
\label{sec:Exercise-simple-diffusion}
Create a Volume Molecule named A with a diffusion constant $10^{-6}\,\text{cm}^2 / \text{s}$, and release a single copy of it at the origin of a Cubic Release Site of diameter zero. Run a simulation for 1 ms (1000 iterations with a time step of $10^{-6}\,\text{s}$), load the visualization data, and play the movie. Note that the molecule may appear very small, so you will probably need to zoom in with the mouse scroll wheel to see it. When you do, you will see two copies of your molecule. The one at the origin is a template that doesn't move.
\index{diffusion}

\section{Restricting diffusion by defining meshes}

\begin{figure}[tbh]
\centering
\includegraphics[width=0.5\textwidth]{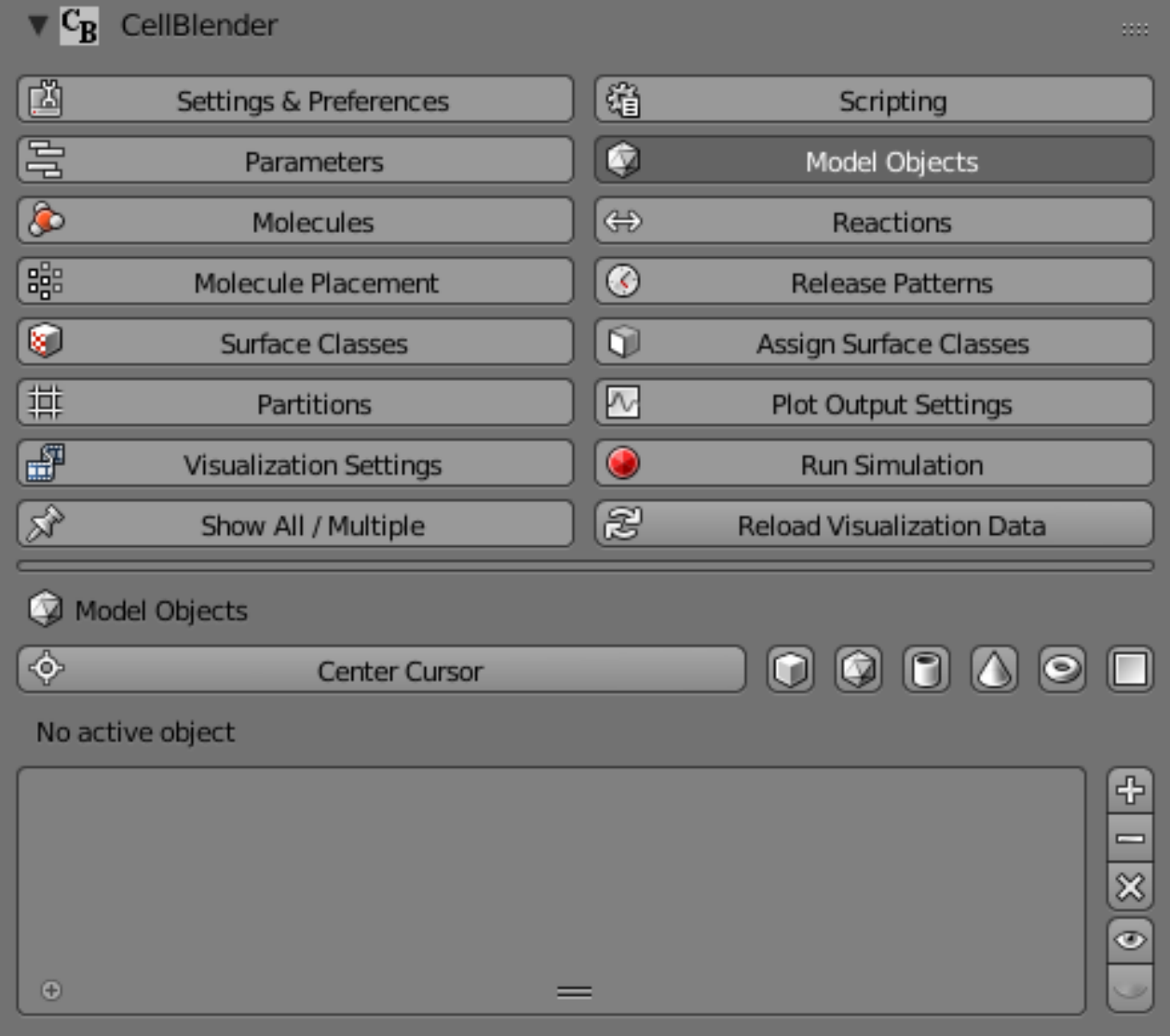}
\caption{ The \textit{Model Objects} panel allows the user to create meshes with a number of different predefined geometries.}
\label{figure:7}
\end{figure}

Model Objects (also called Mesh Objects) may be created using a number of Blender's built-in tools (for example, see \url{http://BlenderArtists.org}). CellBlender provides easy access to a subset of these tools in the \textit{Model Objects} panel (Figure~\ref{figure:7}), which contains controls for centering the cursor, creating primitives (like planes, cubes, and cylinders), renaming objects, and adding objects to the current CellBlender model. In the remaining examples we will use the \textit{Model Objects} panel to create Model Objects that define surfaces on which Surface Molecules diffuse and that can restrict the diffusion of Volume Molecules.

\subsection{CellBlender preliminaries}

\begin{figure}[tbh]
\centering
\includegraphics[width=\textwidth]{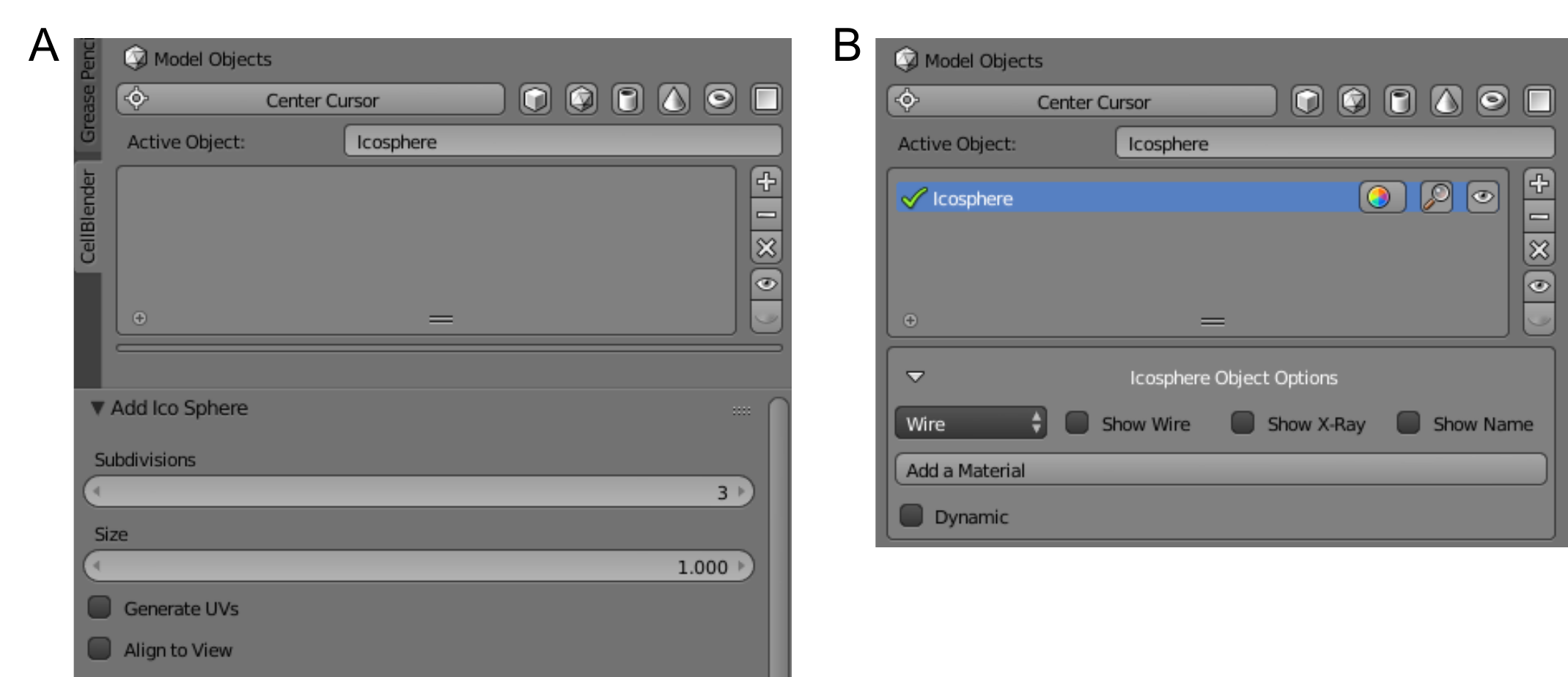}
\caption{
Adding an object to a model.  (A) After clicking one of the object buttons in the \textit{Model Objects} panel, the user can modify various properties of the new Object. In the example shown, the Icosphere has properties Subdivisions and Size. (B) Once the Object is added to the CellBlender model by clicking the ``+'' button, the user may adjust various display properties of the Object.}
\label{figure:8}
\end{figure}

\textbf{Creating Model Objects}. Objects are created at the location of the 3D Cursor (Figure~\ref{figure:4}), which is typically a small circle of alternating red and white segments. Because the 3D Cursor can be tricky to place on the 2D screen, the \textit{Center Cursor} button is provided to make it easier to create CellBlender model objects at the origin. To create a new Model Object, simply place the 3D Cursor at the desired location (the origin for all of the examples in this tutorial), then click one of the object buttons to the right of the \textit{Center Cursor} button---cube, icosphere, cylinder, cone, torus, or plane---to generate a new object at the location of the 3D Cursor. After the object is created, it can be renamed using the \textit{Active Object} field (Figure~\ref{figure:8}(A)). The Object's properties, such as size and number of vertices, may also be modified by changing the numbers entered in the \textit{Add Object} panel that appears (Figure~\ref{figure:8}(A), bottom).

\textbf{Adding Model Objects to a model}. The newly created and named Object must be added to the CellBlender model using the ``+'' button on the right side of \textit{Model Objects}, after which the Object appears in the Object list with a green check mark (Figure~\ref{figure:8}(B)). The Model Objects list also contains controls for changing an object's color and visibility.

\subsection{Exercise: restricted diffusion}

In this exercise you will modify the model from Exercise \ref{sec:Exercise-simple-diffusion} by adding an icosphere object and releasing 1000 A molecules inside that new object. Perform the following steps:
\begin{enumerate}
\item In the \textit{Model Objects} panel center the cursor and then add an icosphere. The default name (``Icosphere'') is fine for this exercise (Figure~\ref{figure:8}(A)). Change the number of  subdivisions from 2 to 3, which gives a smoother icosphere. 
\item With the icosphere is still selected, click the ``+'' button in the Model Objects panel to add the new icosphere to the list of CellBlender model objects. It should show up with the green check mark mentioned earlier. 
\item Open the \textit{Icosphere Object Options} panel directly below the list of Model Objects (Figure~\ref{figure:8}(B)) and change the display settings from Solid to Wire, which renders the Object transparent. As an alternative you can also add materials and give them varying degrees of transparency.
\item In the \textit{Molecule Placement} panel change the number of molecules being released from 1 to 1000, run the simulation with the same time step and duration as before. When the simulation is finished, reload the visualization data again and play the simulation. You should see that all 1000 molecules start out at the origin and are now constrained to stay within the icosphere. 
\item In order to release the 1000 molecules uniformly within the icosphere instead of at a single point, go back to the \textit{Molecule Placement} panel and change the \textit{Release Shape} from \textit{Cubic} to \textit{Object/Region}. Then type the name of the object (``Icosphere'') in the \textit{Object/Region} field below the \textit{Release Shape} selector. If the entry corresponds to a valid Object name, a green check mark will appear in the \textit{Release/Placement Sites} list. Run the simulation again and watch the resulting animation. You should see that the molecules are released uniformly throughout the Icosphere at the start of the simulation.
\end{enumerate}

\section{Simulating bimolecular reactions in a volume}

In this section we show how to include reactions between molecules. We will first consider a simple bimolecular reaction confined within a spherical volume and then describe a more complex system---the Lotka-Volterra model of predator-prey interactions. 
\index{Lotka-Volterra model}

\subsection{CellBlender preliminaries}
\begin{figure}[tbh]
\centering
\includegraphics[width=\textwidth]{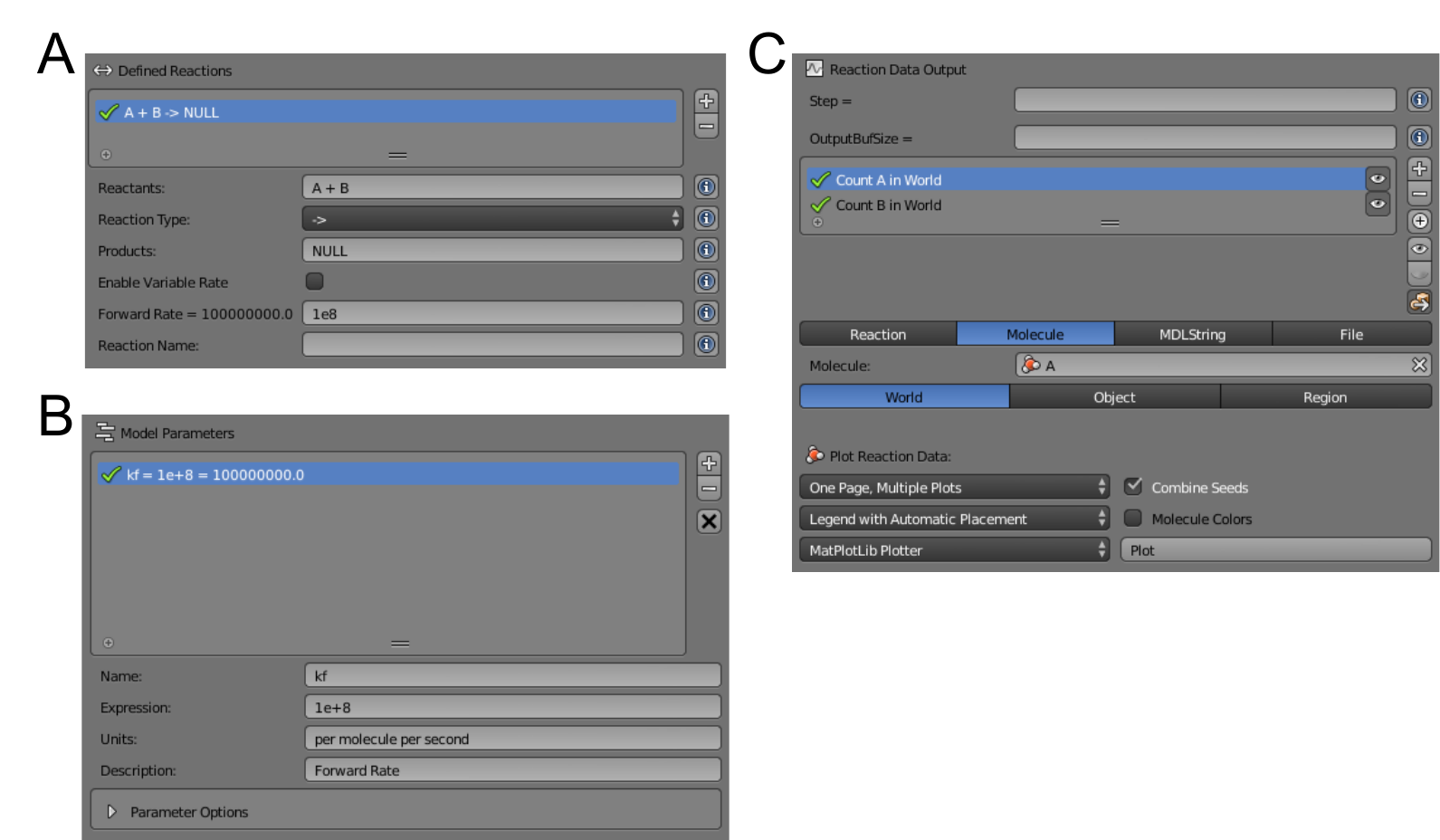}
\caption{Defining Reactions, Parameters, and Plot Output data. (A) \textit{Reactions} panel. (B) \textit{Parameters} panel. (C) \textit{Output Settings} panel. If the variable Step near the top is defined, output will occur at this interval in number of steps; otherwise, output occurs at every time step.}
\label{figure:9}
\end{figure}

This example will make use of the following CellBlender panels:

\textbf{Reactions}. Figure~\ref{figure:9}(A) shows the \textit{Reactions} panel. A Reaction is defined by specifying the reactants separated by plus signs, the type of reaction (irreversible or reversible), the products separated by plus signs, and the rate constants governing the reaction, which may be entered directly as numbers or as expressions that refer to variables defined in the \textit{Parameters} panel. Valid names for reactants are the names of the defined Molecules. CellBlender will report an error if an undefined name is used. In addition, the NULL keyword can be used to indicate the absence of products in a degradation reaction.

\textbf{Parameters}. Figure~\ref{figure:9}(B) shows the \textit{Parameters} panel. Parameters must have a name and a  value, and may further be given Units and a Description. Note that use of Parameters is optional---numerical values may be used instead of named parameters wherever a number is required.

\textbf{Plot Output Settings}. Figure~\ref{figure:9}(C) shows the \textit{Plot Output Settings} panel, which allows the user to define quantities that will be tracked and output to file during a simulation. Both reaction firings and molecule counts can be tracked, and the items included can span the entire model (``World'') or be restricted to a specific Object or Region of an Object. Clicking on one of the Plotter buttons following a simulation will bring up an interactive plot of the specified observables.

\subsection{Exercise: simulating bimolecular degradation}

\begin{figure}[tbh]
\centering
\includegraphics[width=0.5\textwidth]{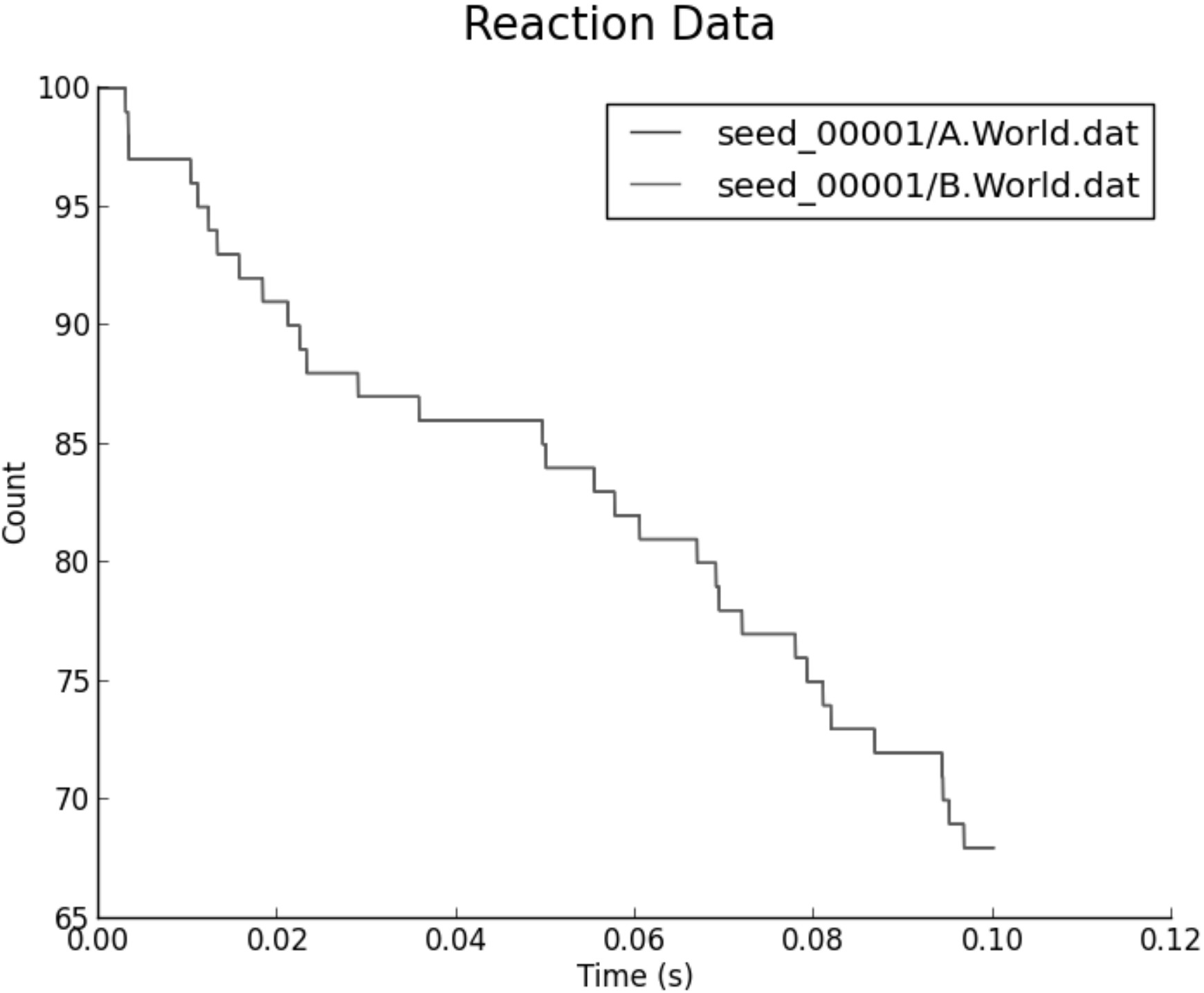}
\caption{Simulation results for a bimolecular reaction plotted using \textit{Simple Plotter}. As A and B collide and annihilate, the counts decrease. Since one molecule of A collides with one molecule of B to annihilate, the graphs for the two species are identical.}
\label{figure:10}
\end{figure}

Define two kinds of Volume Molecules A and B, each with a diffusion constant $10^{-6}\,\text{cm}^2 / \text{s}$. Release 100 copies of each into an icosphere, and simulate the bimolecular degradation reaction {A+B$\rightarrow$NULL} with a rate constant of $10^8\,\text{M}^{-1} \text{s}^{-1}$. Define output observables to count the number of A and B molecules. Run your simulation for $10^4$ iterations with a time step of  $10^{-5}\,\text{s}$ and plot the results, which should resemble those shown in Figure~\ref{figure:10}. 

\subsection{Exercise: The Lotka-Volterra predator-prey model}

The Lotka-Volterra equations describe the interactions of two species, one of which preys upon the other. Although usually analyzed in the well-mixed context, this model can exhibit strong spatial effects under some conditions, as we will see by developing an MCell model of this system.
\index{Lotka-Volterra model}

The basic interactions can be described with the following set of reactions, where the prey are sheep (S) and predators are wolves (W):
\begin{gather*}
\mathrm{S \rightarrow S + S} \\
\mathrm{S + W \rightarrow W + W} \\
\mathrm{W \rightarrow NULL},
\end{gather*} 
which can be summarized as: prey can spontaneously multiply, predators can eat prey to multiply, and predators spontaneously die. In our treatment, the species in the system will be confined to a thin slab, which could represent a two-dimensional forest.

\subsubsection{Reaction-limited behavior}
\begin{figure}[tbh]
\centering
\includegraphics[width=\textwidth]{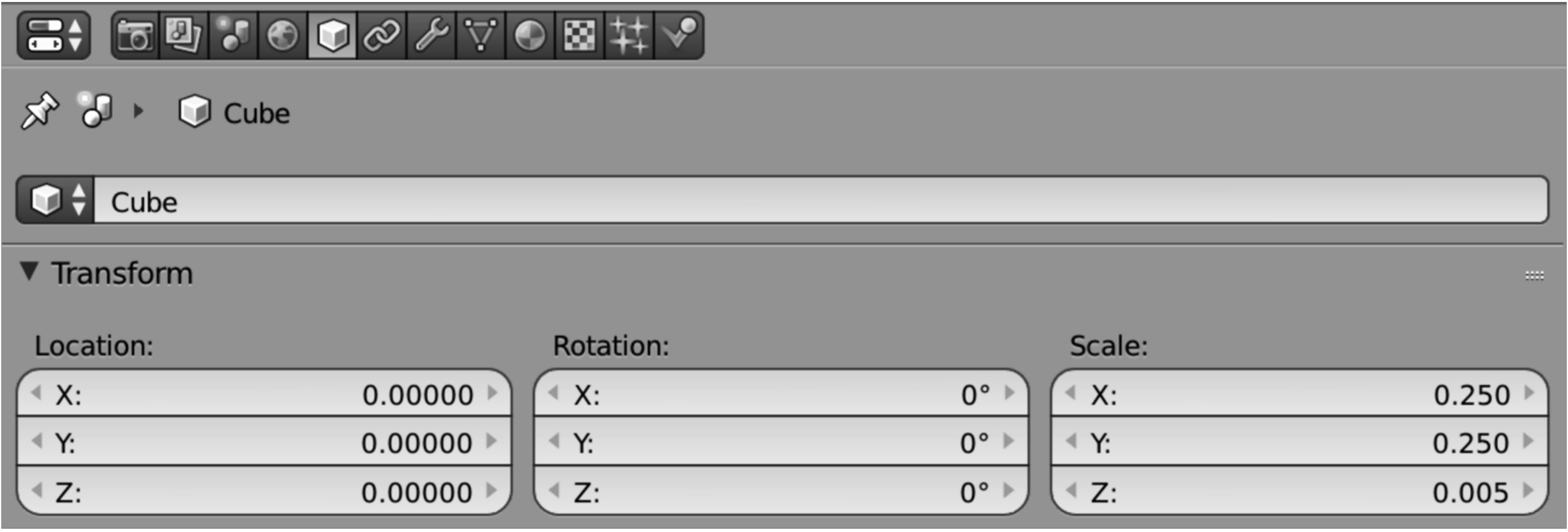}
\caption{Changing Object properties in Blender. Selecting the cube icon in the bar at the top of the \textit{Properties} panel (right hand column of main Blender window shown in Figure~\ref{figure:4}) brings up a list of panels with options for changing properties of the selected Object. The \textit{Transform} panel enables translation, rotation, and scaling. Changing the X and Y scales to 0.25, and the Z scale of a cube to 0.005 creates a thin slab for the Lotka-Volterra model simulations. Set the display type to \textit{Bounds} in \textit{Object Options} as you did in the previous example to make the box transparent. Then use the middle mouse button (two-finger swipe on Mac) to rotate the view to view from the top of the slab, and then zoom with the right mouse button (two-finger pinch on Mac) to fill the window.}
\label{figure:11}
\end{figure}

From the \textit{Model Objects} panel, create a cube centered at the origin. Use Blender to deform this cube into a thin slab by adjusting the scales for the X, Y and Z axes as shown in Figure~\ref{figure:11}, and add the cube to your model. Define S and W as Volume Molecules with diffusion constant  $6\times10^{-6}\,\text{cm}^2 / \text{s}$. Release 1000 copies of each into the thin slab by defining an appropriate Release Site of type Object/Region. Enter the reactions defined above, with the rate constants $1.29\times10^{5}\,\text{s}^{-1}$, $10^{8}\,\text{M}^{-1}\,\text{s}^{-1}$, and $1.30\times10^{5}\,\text{s}^{-1}$ respectively. Define the \textit{Plot Output Settings} to track the total counts of S and W and run the simulation for 500 iterations with a time step of $10^{-6}\,\text{s}$. After the run completes, reload the visualization data and plot the results.
\index{reaction-limited reactions}

\begin{figure}[tbh]
\centering
\includegraphics[width=0.75\textwidth]{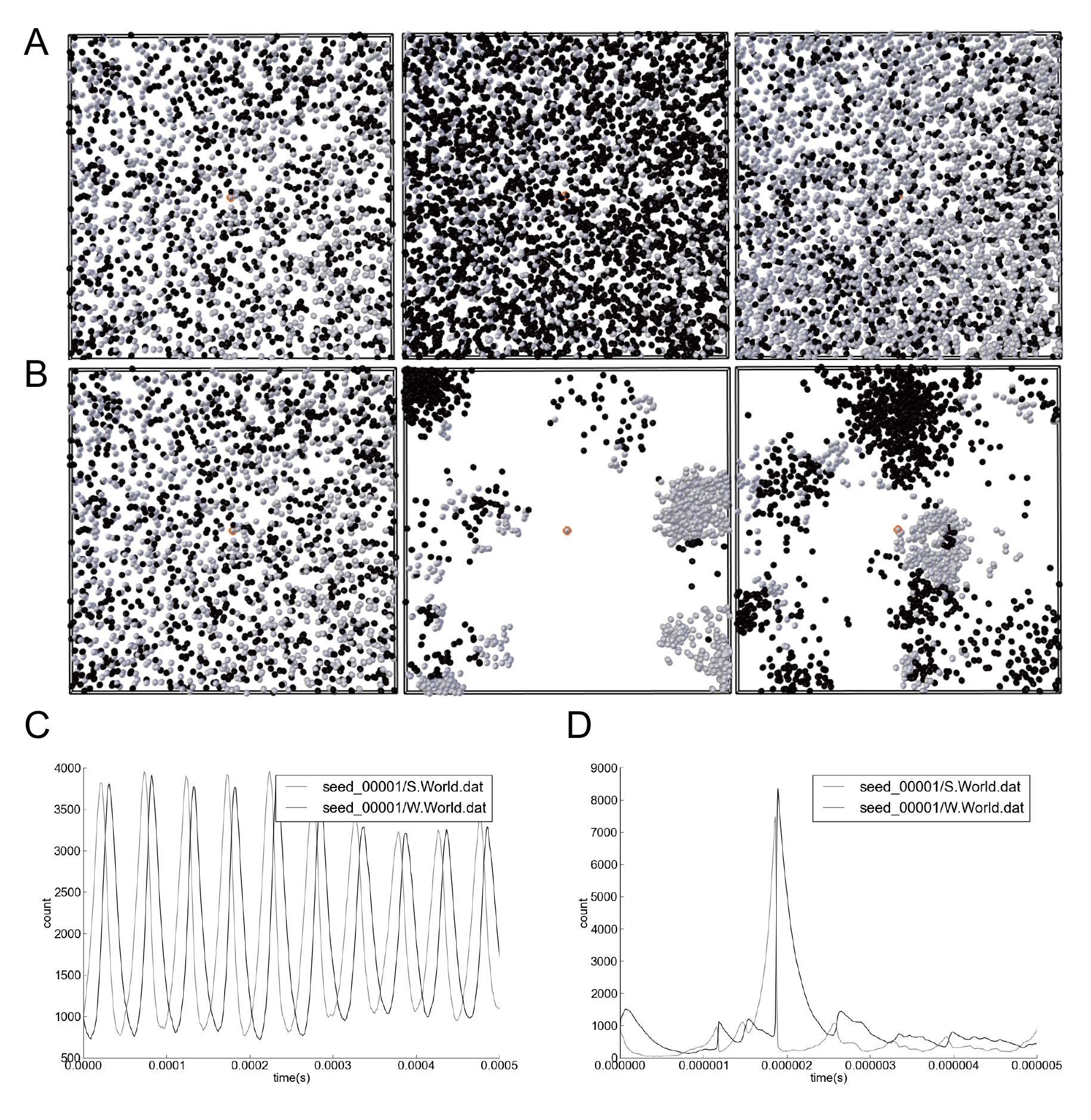}
\caption{Sheep and wolf populations in the Lotka-Volterra model in the reaction-limited (A,C) and diffusion-limited (B,D) regimes. (A) Simulation snapshots along a trajectory in the reaction limited regime at 0, 140, and 320 ms going from left to right. (B) Simulation snapshots along a trajectory in the diffusion limited regime at 0, 140, and 280 ms. (C,D) Populations of S (light lines) and W (dark lines) versus time for the reaction-limited (C) and diffusion-limited (D) regimes.}
\label{figure:12}
\end{figure}

For these parameters the system is \emph{reaction-limited}, because diffusion is fast compared with the rates of reaction, meaning that the concentrations are relatively uniform in space but not in time, which is the behavior observed in  Figure~\ref{figure:12}(A,C). The S and W concentrations vary in time but remain spatially well-mixed. The time courses shown in Figure~\ref{figure:12}(C) would be very similar to trajectories obtained by solving the corresponding ODE's or using Gillespie's algorithm.

\subsubsection{Diffusion-limited behavior}

Now increase the reaction rates to $8.6\times10^{6}\,\text{s}^{-1}$, $10^{12}\,\text{M}^{-1}\,\text{s}^{-1}$, and $5.0\times10^{6}\,\text{s}^{-1}$ respectively.
Reduce the time step to $10^{-8}\,\text{s}$ and run a simulation again for 500 iterations. 
The reduced time step ensures that the probabilities of unimolecular reactions stay below one, which is necessary to maintain accuracy. (You can check this by running the simulation again with the \textit{Save Text Logs} option enabled in the \textit{Output/Control Options} subpanel of the \textit{Run Simulation} panel. After running the simulation, click on the \textit{Screen Layout} button, which is next to \textit{Help} on the menu bar at the top of the CellBlender interface (Figure~\ref{figure:4}), and select the \textit{Scripting} layout. This layout allows you to see the MCell text log, which reports the reaction probabilities for each reaction with the specified time step. The output log for each simulation run is stored in a separate file with the name task\_\textit{PID}\_output, where \textit{PID} is the process ID reported in the \textit{Run Simulation} panel. To view the log, click the notebook icon to the left of the \textit{New} button and select the log name from the drop-down menu.) The reaction probability for the one bimolecular reaction remains above one, which is acceptable in this case because it simply means that every collision results in a reaction. This is the expected behavior in the diffusion limit.
\index{diffusion-limited reactions}

For these reaction rate constants, the system becomes \emph{diffusion-limited}, because diffusion becomes slow relative to the rates of reaction. Because of the strong predator-prey interaction, S molecules in the vicinity of a W molecule are rapidly converted to W, and the S and W populations tend to segregate. The overall system exhibits dramatic spatial heterogeneity with irregular oscillations (Figure~\ref{figure:12}(B,D)).

\section{Simulating molecules and reactions on surfaces}

So far we have considered models in which molecules move and react in a 3D volume. Many molecules involved in cell signaling, however, are confined to membrane surfaces on and within cells. Extracellular ligands binding to cell surface receptors are a principal means by which cells sense their environment. Molecules confined to membranes have higher effective concentrations and thus membrane localization enhances catalysis and binding among surface-associated molecules, which is a critical mechanism for cell signaling \cite{Haugh1997,Kholodenko2000,Haugh2002}. This section provides an introduction to modeling molecules and reactions on surfaces in MCell.
\index{cell signaling}

\subsection{CellBlender preliminaries} 

\begin{figure}[tbh]
\centering
\includegraphics[width=\textwidth]{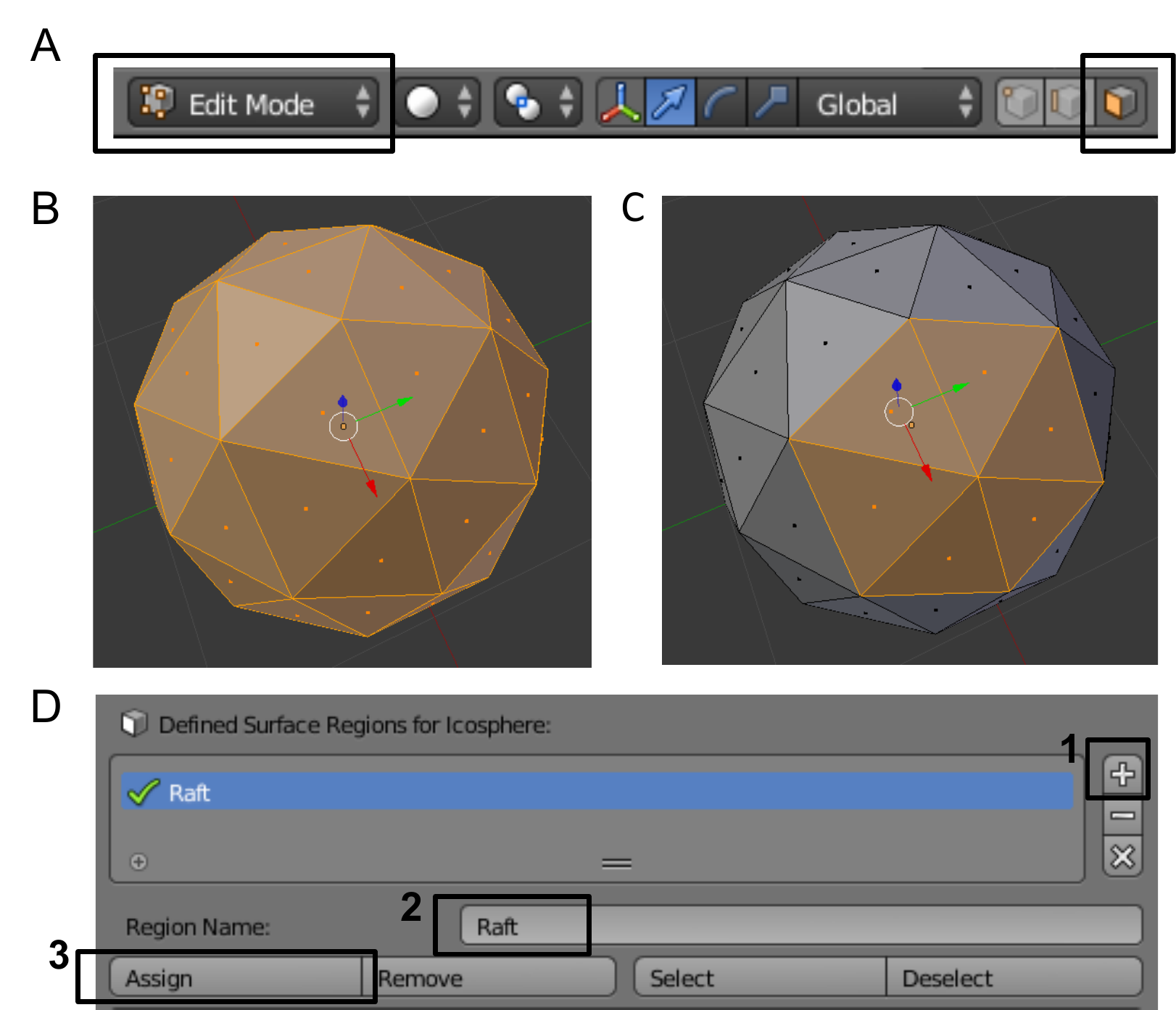}
\caption{Defining Surface Regions. (A) The menu bar near the bottom of the Blender interface (Figure~\ref{figure:4}) with options to enter \textit{Edit Mode} (left) and use \textit{Face Select} (right). (B) Triangulated Object with all Faces selected. (C) Triangulated Object with six Faces selected. (D) \textit{Defined Surface Regions} subpanel at the bottom of the \textit{Model Objects} panel showing the steps required to define a Surface Region from a set of selected Faces.}
\label{figure:13}
\end{figure}

\begin{figure}[tbh]
\centering
\includegraphics[width=\textwidth]{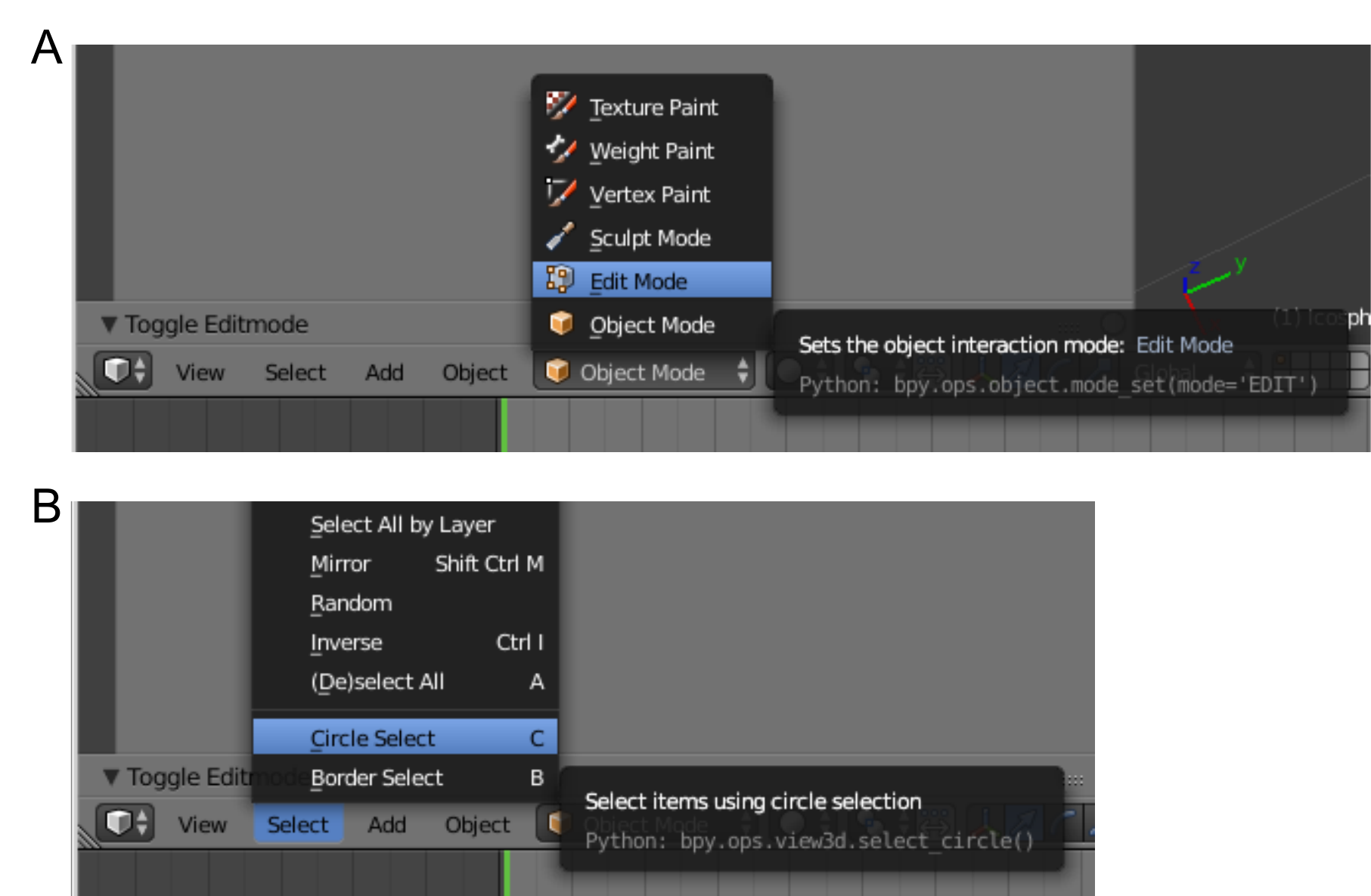}
\caption{Selecting Faces using \textit{Circle Select}. (A) Object interaction model menu. \textit{Edit Mode} is used to select Faces of an object to define Surface Regions. (B) The \textit{Circle Select} option can be used to select Faces on a triangulated Object. Faces are selected by holding the left mouse button while dragging the cursor over the desired Faces.}
\label{figure:14}
\end{figure}

Surfaces in MCell have a distinct orientation (Front/Back) that is determined by the direction of a surface normal determined from the right-hand rule by the order of vertices in each triangle. To be a valid surface in MCell, the ordering of vertices in each triangle must give a consistent direction for the surface normal; fortunately, CellBlender takes care of this automatically. Molecules on surfaces also have an orientation (Top/Bottom) that can be aligned with or against the surface normal. The orientation of surfaces and molecules becomes important in specifying placement of Surface Molecules and in specifying reactions that involve either Surfaces or Surface Molecules. Syntactic features of reactions that involve surfaces are presented below. Here we briefly describe the elements required to specify surfaces and surface reactions:

\textbf{Surface Regions}. Surface Regions are composed of a subset of the Faces making up an Object. They are used to control the placement of Surface Molecules and to allow variation in surface properties in different regions---in particular how Surface Molecules move and how Volume Molecules interact with surfaces. Figure~\ref{figure:13} shows the steps involved in creating a Surface Region for an existing CellBlender Model Object. First, with the Object selected the user enters Blender's \textit{Edit Mode} by hitting the Tab key with the cursor in the main window or by selecting \textit{Edit Mode} in the menu bar near the bottom of the Blender window (Figure~\ref{figure:13}(A)). Clicking the \textit{FaceSelect} button to the right on the same bar will enable \textit{Face Select Mode} (Figure~\ref{figure:13}(B)). Now, the user can select a set of Faces by right-clicking on the first Face and then holding Shift while right clicking to select additional Faces (Figure~\ref{figure:13}(C)). Figure~\ref{figure:14} presents an alternate way to select Faces using the \textit{Circle Select Tool}, which is more efficient for large Objects. The \textit{Defined Surface Regions} subpanel, which is part of the \textit{Model Objects} panel, is then used to create a named Surface Region containing the selected Faces (Figure~\ref{figure:13}(D)). The \textit{Select} and \textit{Deselect} buttons can be used to toggle the selected Faces and to determine which Faces are assigned to a Surface Region.  Note that it is allowable for Surface Regions to overlap.

\begin{figure}[tbh]
\centering
\includegraphics[width=\textwidth]{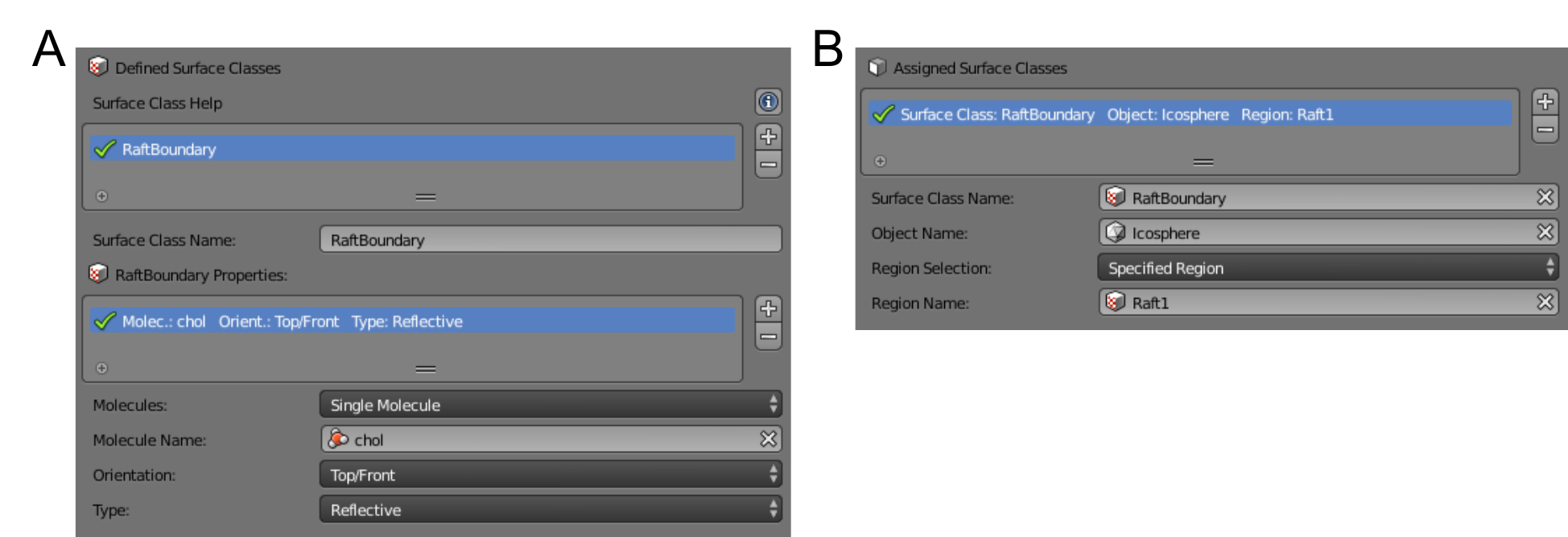}
\caption{Defining and Assigning Surface Classes. (A) The \textit{Surface Classes} panel. Each class (top list) may have multiple properties (bottom list). (B) The \textit{Assign Surface Classes} panel. Both Object Name and Region Name are required.}
\label{figure:15}
\end{figure}

\textbf{Surface Classes}. Surface Classes define how a Molecule behaves when it encounters a surface, e.g., whether it reflects, passes through, or is absorbed. They allow for development of realistic models because biological membranes may have different properties with respect to different molecules. For example, some molecules may freely diffuse across a membrane, whereas others may not. It is also useful to define ``transparent'' membranes for the purpose of counting particular types of molecules in particular regions of space. Figure~\ref{figure:15}(A) shows the CellBlender interface for defining a Surface Class, which requires a name and one or more Properties and which can be of various types including  \textit{Transparent}, \textit{Reflective}, and \textit{Absorptive}. Each Property applies to only one type of Molecule (and may further depend on Orientation), and many Properties can be defined for a single Surface Class. For any Molecule--Surface interaction that is not governed by a defined Surface Class, a default behavior is assumed: \textit{Reflective} for Volume molecules and \textit{Transparent} for Surface Molecules.

\textbf{Assigning Surface Classes}. Surface Classes are assigned to Surface Regions in the \textit{Assign Surface Classes} panel (Figure~\ref{figure:15}(B)). Note that to apply a Surface Class to a whole Object, a Surface Region including all Faces in the Object must be defined. 

\textbf{Defining surface reactions}. 
Surface reactions in MCell may be defined with reference to either the relative or absolute orientation of the reacting molecules, as introduced in \cite{Kerr2008}. For example, to specify a reaction where a Surface Molecule B flips its orientation, we write 
\[
\mathrm{B'\rightarrow B,}
\]
where the apostrophe and comma represent the relative orientation of B on the reactant and product side respectively. Note that whether the tick mark is up or down on the reactant side does not matter; what matters is that the position changes. MCell uses a shorthand notation where the number of ticks indicates the orientation class of a surface object, while the position (up or down) indicates the relative orientation. Thus, the reaction 
\[
\mathrm{B,\rightarrow B'}
\]
is equivalent to the previous one: it will flip the relative orientation of any B molecule on a surface regardless of its absolute orientation on that surface. Any reaction that involves a Surface Molecule must specify the relative orientations of all reactants and products. In this example
\[
\mathrm{B'\rightarrow B' + A' + C},
\]
B retains its orientation while A and C, if they are Surface Molecules, are generated with the same and opposite orientations respectively because they are in the same orientation class as the reactant B. If instead A and C are Volume Molecules, then the orientation class indicates the side of the surface the products will be generated on. Here, A would be generated on the side where B is pointing and C would be generated on the opposite side. If all of the B molecules have been given the same orientation, then A and C will always be produced on opposite sides of the surface. The absolute polarity of this model would be maintained as long as there are no reactions that flip the orientation of B. 

It is sometimes useful to specify absolute orientation in reactions, which requires adding a Surface Class to the specification of reactants.  For example, consider the following reaction where L is a Volume Molecule and R and LR are Surface Molecules:
\[
\mathrm{L' + R' @ surf' \rightarrow LR'}
\]
Here the fact that the Surface Molecule R and the Surface Class \textbf{surf} are in the same orientation class requires that R have a fixed absolute orientation (Top-Front) with respect to \textbf{surf}. In addition, L is required to be in a volume adjacent to the Front of \textbf{surf}, and LR will be produced with the same orientation as R. For a more details about surface reactions, see \url{http://mcell.org/documentation}.

\subsection{Exercise: a simple model of receptors aggregation in lipid rafts}

\begin{figure}[tbh]
\centering
\includegraphics[width=\textwidth]{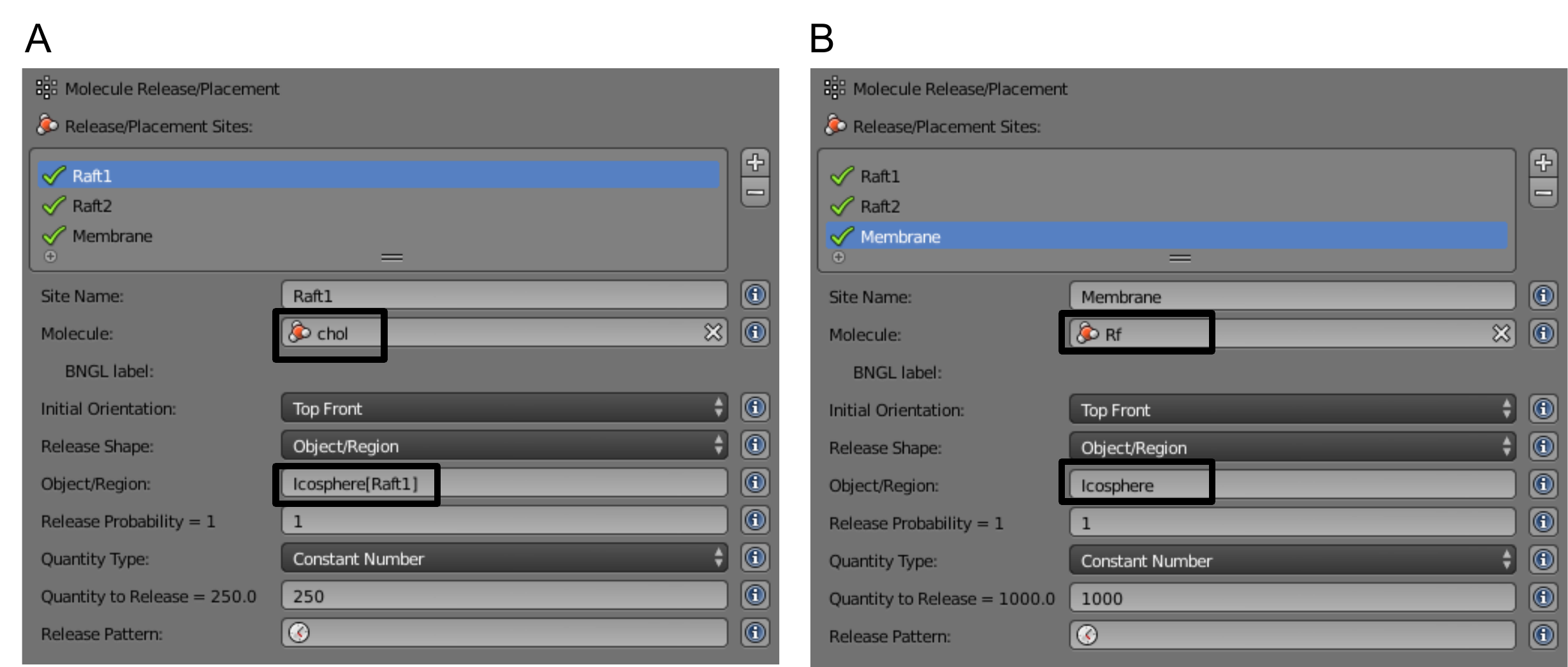}
\caption{Releasing Molecules on Surface Regions. (A) The Surface Region is specified with the syntax  ObjectName[RegionName] in the \textit{Object/Region field}. (B) Specifying only the Object name causes Molecule release over the full surface.}
\label{figure:16}
\end{figure}

\begin{figure}[tbh]
\centering
\includegraphics[width=\textwidth]{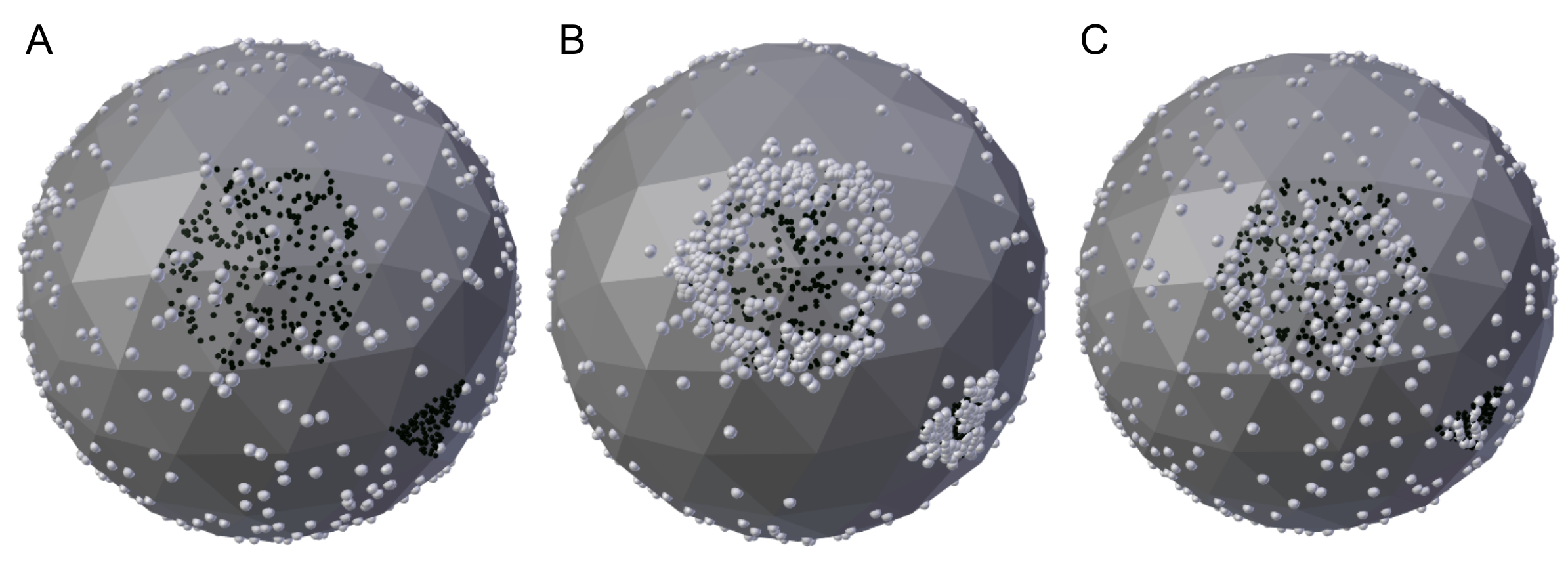}
\caption{Simulation results for the model of receptor aggregation in lipid rafts. (A) Initial configuration with chol (black circles) confined to rafts and uniform distribution of receptors (white spheres). (B) Snapshot at 1 $\mu$s showing receptor clustering in the two raft regions. (C) Snapshot at 1 $\mu$s shown reduced clustering due to the addition of a reaction that converts Rs back to Rf.}
\label{figure:17}
\end{figure}

We will create a simple model of receptor clustering in lipid rafts. Fast-diffusing receptor molecules (Rf) react with cholesterol molecules (chol) in rafts to become slow-diffusing receptors (Rs). The model can be built in the following steps:
\begin{enumerate}
\item Create an icosphere to represent a cell and add it to your model. Increase the number of subdivisions to 3 to make the sphere smoother. 
\item Create two non-touching Surface Regions named Raft1 and Raft2 comprised of 6 triangles forming a hexagon for Raft1 and one triangle for Raft2 following the procedures shown in Figures~\ref{figure:13} and \ref{figure:14}. 
\item Define Surface Molecules Rf, Rs, and chol with diffusion constants $10^{-6}\,\text{cm}^2 / \text{s}$, $10^{-9}\,\text{cm}^2 / \text{s}$, and $10^{-6}\,\text{cm}^2 / \text{s}$ respectively.
\item Create a Surface Class that is reflective with respect to chol (Figure~\ref{figure:15}) and assign it to both Surface Regions (Figure~\ref{figure:16}). 
\item Release 250 chol molecules in Raft1 and 100 chol molecules in Raft2 as shown in Figure~\ref{figure:16}(A) and 1000 Rf molecules distributed over the whole surface as shown in Figure~\ref{figure:16}(B).
\item Define the surface reaction
\[
\mathrm{Rf' + chol' \rightarrow Rs' + chol'}
\]
and assign it a rate constant of $10^{8}\,\mu\text{m}^2 / \text{s}$.
\item Run your simulation for 1000 iterations at a time step of $10^{-5} \text{s}$. If you change the colors in \textit{Display Options} of the Molecules panel so that Rs and Rf are the same color, your results should resemble those shown in Figure~\ref{figure:17}(A,B). Because there is no reaction to convert Rs back to Rf, Rs molecules accumulate at the raft boundaries, particularly Raft1 (Figure~\ref{figure:17}(B)). Eventually all receptors will end up as Rs. 
\item To make the model more realistic, modify the previous reaction scheme to include a reverse reaction:
\[
\mathrm{Rs' \rightarrow Rf' }
\]
and give it a rate constant $10^4 \,\text{s}^{-1}$. 
Simulating the extended model should give results similar to those shown in Figure~\ref{figure:17}(C). There is still pronounced clustering in the rafts, but the receptors are more uniformly distributed throughout each raft.
\end{enumerate}
\index{lipid raft}
\index{receptor clustering}

\section{Extended exercise: a density-dependent switch}

We now consider a second mechanism for receptor clustering based on positive feedback \cite{Jilkine2011}. In this model receptors can spontaneously become activated at a slow rate, but upon activation can activate additional receptors at a faster rate, which represents a form of positive feedback. The strength of the feedback depends on the receptor density, which in turn affects the spatial heterogeneity, as we will observe. Since all of the CellBlender operations needed to specify this model were presented in the previous sections, we can jump into a description of the model. 
\index{positive feedback}
\index{receptor clustering}

Create Surface Molecules R and Ra, which represent the inactive and active forms of the receptor, and set the diffusion constants to be $10^{-7}\,\text{cm}^2\,\text{s}^{-1}$ and $10^{-10}\,\text{cm}^2\,\text{s}^{-1}$ respectively. The reaction scheme is 
\begin{gather*}
\mathrm{R' \rightarrow Ra'} \\
\mathrm{Ra' + R' \rightarrow Ra' + Ra'} \\
\mathrm{Ra' \rightarrow R'} 
\end{gather*}
The rate constants for the three reactions are $10^{-5}\,\text{s}^{-1}$, $40\,\mu\text{m}^2 \,\text{s}^{-1}$, and $40\,\text{s}^{-1}$ respectively.

\begin{figure}[tbh]
\centering
\includegraphics[width=\textwidth]{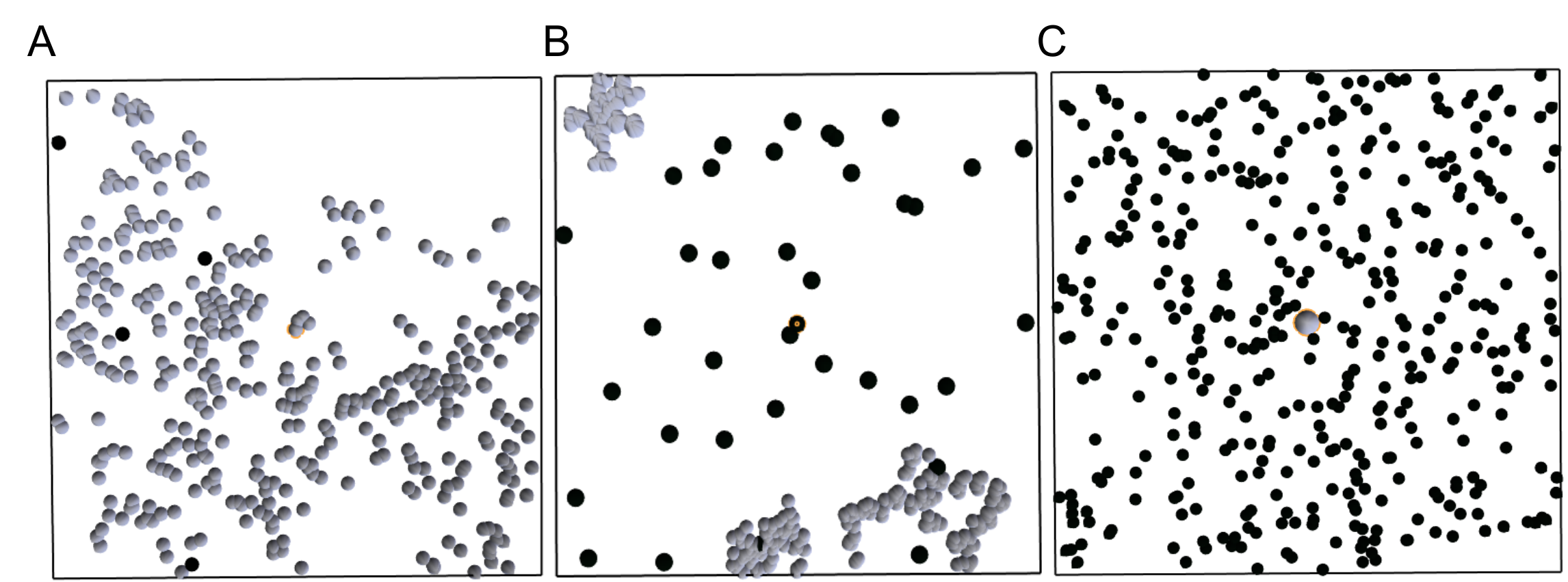}
\caption{Simulation snapshots after 1 s simulation for the density dependent switch at (A) high density, (B) medium density, and (C) low density showing active (gray spheres) and inactive receptors (black circles).  Model parameters are given in the text.}
\label{figure:18}
\end{figure}

\textbf{High density}. Use the \textit{Model Objects} panel to create a Plane Object (shape furthest to the right) with surface area  $1\,\mu\text{m}^2$ by setting Radius to 0.5 in the \textit{Add Plane} subpanel. Add this Object to your model. Release 5 molecules of Ra and 395 molecules of R on to the plane. Define the reaction scheme and parameters as mentioned above, and run your simulation for $5,000$ iterations with a time step of $10^{-5}\,\text{s}$. Play your simulation and compare with the results shown in Figure~\ref{figure:18}(A). At this density the active state is stabilized with respect to the inactive state and nearly all of the receptors end up as active. Because the active receptors diffuse slowly on the time scale of this simulation, clusters arise around the initially active receptors, and then these clusters merge as the proportion of active receptors grows further.

\textbf{Medium density}. Increase the area of the plane from $1\,\mu\text{m}^2$  to $16\,\mu\text{m}^2$ using the \textit{Transform} panel as described above (Figure~\ref{figure:11}). For how many iterations do you need to simulate now in order to approach the steady state for number of active receptors? Compare your results to those shown in Figure~\ref{figure:18}(B). At this lower density, clusters persist and remain distinct unless the seeding Ra molecules were initially close.

\textbf{Low density}. Increase the surface area again to $1600\,\mu\text{m}^2$  and rerun your simulation ($10,000$ iterations should suffice). At this density the rate of receptor-catalyzed activation is too low to maintain receptor activity, and all of the Ra molecules eventually convert to R molecules (Figure~\ref{figure:18}(C)). 

Changing the area to vary the system density shows that the system can exhibit at least three distinct behaviors going from high to low density (Figure~\ref{figure:18}): global clustering that links distinct sites of receptor activation, local clustering, in which each site of spontaneous activation nucleates a distinct cluster, and extinction, in which activation of a receptor results in little or no activation of nearby receptors. Such effects may be important for signal propagation and would be difficult to capture using a modeling framework that does not capture both spatial and stochastic effects.

\section{Concluding remarks}

Models play a valuable role in biology by both aiding our understanding of old observations and predicting new ones. As discussed in the Introduction, modeling formalisms vary in their molecular and spatial resolution, and understanding the limitations of each will help a modeler choose an appropriate method for a given problem. This chapter has introduced the theory and algorithms that can be used to carry out one form of spatial stochastic modeling as implemented in the MCell simulation package. It has also provided a tutorial for building, simulating, and analyzing MCell models using the CellBlender interface starting from basic diffusion and reaction-diffusion systems, and building up to more advanced examples that demonstrate how spatial heterogeneity can arise from a variety of mechanisms including diffusion-limited reactions (Lotka-Volterra), membrane organization and hindered diffusion (lipid rafts), and positive feedback. 

For more information about MCell and CellBlender visit \newline \url{http://www.mcell.org/}.

\section*{Acknowledgment}
We would like to thank Markus Dittrich for his many contributions to MCell and CellBlender and the participants of the 2016 Cell Modeling Workshop and the 2016 q-bio Summer School (University of New Mexico campus) for their helpful input on the examples. This work was supported in part by the NIGMS-funded (P41-GM103712) National Center for Multiscale Modeling of Biological Systems (MMBioS) and NIH grant R35-GM119462.

\end{document}